\documentclass[acmsmall,screen]{acmart}

\setcopyright{rightsretained}
\copyrightyear{2020}
\acmJournal{PACMPL}
\acmYear{2020} \acmVolume{4} \acmNumber{POPL} \acmArticle{51} \acmMonth{1}
\acmPrice{}\acmDOI{10.1145/3371119}

\ccsdesc[500]{Software and its engineering~Software libraries and repositories}
\ccsdesc[300]{Theory of computation~Logic and verification}
\ccsdesc[300]{Theory of computation~Equational logic and rewriting}
\ccsdesc[500]{Theory of computation~Denotational semantics}

\keywords{Coq, monads, coinduction, compiler correctness}

\newif\ifcomments\commentsfalse   
\newif\ifaftersubmission \aftersubmissionfalse 
\newif\ifplentyofspace \plentyofspacefalse 
\newif\ifredacted \redactedtrue 

\IfFileExists{texdirectives}{\input{texdirectives}}{}

\ifredacted
\newcommand\redacted[1]{[REDACTED]}
\else
\newcommand\redacted[1]{#1}
\fi

\makeatletter
   \@printpermissiontrue
   \@printcopyrighttrue
   \@acmownedtrue
\makeatother
\def\authorsaddresses#1{}

\usepackage{booktabs}   
\usepackage{subcaption} 
\usepackage{cleveref}

\usepackage{listings,lstlangcoq}
\usepackage{stmaryrd}
\usepackage{bussproofs}

\usepackage[export]{adjustbox}
\usepackage[normalem]{ulem}
\usepackage{balance}
\usepackage{tikz}
\usetikzlibrary{calc}
\usepackage{caption,subcaption}
\usepackage[inline]{enumitem}

\lstdefinestyle{customc}{
  belowcaptionskip=1\baselineskip,
  breaklines=true,
  xleftmargin=\parindent,
  language=C,
  showstringspaces=false,
  basicstyle=\ttfamily,
  keywordstyle=\bfseries\color{green!40!black},
  commentstyle=\itshape\color{purple!40!black},
  identifierstyle=\color{blue!50!black},
  stringstyle=\color{orange},
}

\lstdefinestyle{customcoq}{
  columns=flexible,
  mathescape=true,
  belowcaptionskip=1\baselineskip,
  breaklines=true,
  xleftmargin=\parindent,
  language=Coq,
  morekeywords={Variant, fun, Arguments, Type, cofix},
  emph={%
    SOCKAPI,ITree,data_at,data_at_
  },
  emphstyle={\bfseries\color{green!40!red!80}},
  showstringspaces=false,
  basicstyle=\small\ttfamily,
  keywordstyle=\bfseries\color{green!20!black},
  commentstyle=\itshape\color{red!40!black},
  identifierstyle=\color{violet!50!black},
  stringstyle=\color{orange},
  escapeinside={<@}{@>}
}

\lstdefinestyle{customocaml}{
  columns=flexible,
  mathescape=true,
  belowcaptionskip=1\baselineskip,
  breaklines=true,
  xleftmargin=\parindent,
  language=caml,
  emphstyle={\bfseries\color{green!40!red!80}},
  showstringspaces=false,
  basicstyle=\small\ttfamily,
  keywordstyle=\bfseries\color{green!20!black},
  commentstyle=\itshape\color{red!40!black},
  identifierstyle=\color{violet!50!black},
  stringstyle=\color{orange},
  escapeinside={<@}{@>}
}

\newcommand{\namedsection}[1]{\lstinputlisting[
        style=customcoq,
        basicstyle=\footnotesize\ttfamily,
        rangebeginprefix=(*\ begin\ ,%
        rangebeginsuffix=\ *),%
        rangeendprefix=(*\ end\ ,%
        rangeendsuffix=\ *),
        includerangemarker=false,
        linerange={#1},]{./ICFPPaper.v}}

\newcommand{\snamedsection}[1]{\lstinputlisting[
        style=customcoq,
        basicstyle=\small\ttfamily,
        rangebeginprefix=(*\ begin\ ,%
        rangebeginsuffix=\ *),%
        rangeendprefix=(*\ end\ ,%
        rangeendsuffix=\ *),
        includerangemarker=false,
        linerange={#1},]{./ICFPPaper.v}}

\ifcomments
\newcommand{\proposecut}[1]{\ifcomments\sout{#1}\fi}
\newcommand{\lx}[1]{\textcolor{olive}{{[LX:~#1]}}}
\newcommand{\yz}[1]{\textcolor{blue}{{[YZ:~#1]}}}
\newcommand{\bcp}[1]{\textcolor{violet}{{[BCP:~#1]}}}
\newcommand{\lys}[1]{\textcolor{green!50!black!100}{{[LYS:~#1]}}}
\newcommand{\nk}[1]{\textcolor{blue!50!green!100}{{[NK:~#1]}}}
\newcommand{\wm}[1]{\textcolor{orange}{{[WM:~#1]}}}
\newcommand{\wh}[1]{\textcolor{red!75!black!100}{{[WH:~#1]}}}
\newcommand{\sz}[1]{\textcolor{brown!100!black!100}{{[SZ:~#1]}}}
\newcommand{\gmm}[1]{\textcolor{blue!100!black!100}{{[GM:~#1]}}}
\newcommand{\lb}[1]{\textcolor{cyan!100!black!100}{{[LB:~#1]}}}
\newcommand{\ph}[1]{\textcolor{orange}{{[PH:~#1]}}}
\newcommand{\gh}[1]{\textcolor{white!50!blue!100}{{[GH:~#1]}}}
\else
\newcommand{\proposecut}[1]{}
\newcommand{\ly}[1]{}
\newcommand{\yz}[1]{}
\newcommand{\lx}[1]{}
\newcommand{\bcp}[1]{}
\newcommand{\lys}[1]{}
\newcommand{\nk}[1]{}
\newcommand{\wm}[1]{}
\newcommand{\wh}[1]{}
\newcommand{\sz}[1]{}
\newcommand{\gmm}[1]{}
\newcommand{\lb}[1]{}
\newcommand{\ph}[1]{}
\newcommand{\gh}[1]{}
\fi
\newcommand{\todo}[1]{\textcolor{red}{{[TODO:~#1]}}}

\newcommand{\op}{\mathtt{op}}

\newcommand{\inlinecoq}[1]{\mbox{\lstinline[style=customcoq,columns=fixed,basewidth=.48em]{#1}}}
\newcommand{\inlinecoqt}[1]{\mbox{\lstinline[style=customcoq,basicstyle=\footnotesize\ttfamily,columns=fixed,basewidth=.48em]{#1}}}

\newcommand{\ilc}[1]{\inlinecoq{#1}}
\newcommand{\ilct}[1]{\inlinecoqt{#1}}

\newcommand{\eqktree}{\hat{\approx}}
\newcommand{\PROP}{\ensuremath{\mathbb{P}}}

\newcommand*{\Imp}{\textsc{Imp}}
\newcommand*{\Asm}{\textsc{Asm}}

\newtheorem{theorem}{Theorem}
\newtheoremstyle{invisible}{}{}{\itshape}{}{}{}{0pt}{}
\theoremstyle{invisible}
\newtheorem*{invisible}{}

\begin{document}


\title{Interaction Trees}
\subtitle{Representing Recursive and Impure Programs in Coq}


\newcommand\pennauthor[1]{#1\textsuperscript\textdagger}
\newcommand\etc{\textit{etc.}}

\author[L. Xia]{Li-yao Xia}
\affiliation{
  \institution{University of Pennsylvania}
  \city{Philadelphia}\state{PA}
  \country{USA}}
\author[Y. Zakowski]{Yannick Zakowski}
\affiliation{
  \institution{University of Pennsylvania}
  \city{Philadelphia}\state{PA}
  \country{USA}}
\author[P. He]{Paul He}
\affiliation{
  \institution{University of Pennsylvania}
  \city{Philadelphia}\state{PA}
  \country{USA}}
\author[C. Hur]{Chung-Kil Hur}
\affiliation{
  \institution{Seoul National University}
  \city{Seoul}
  \country{Republic of Korea}}
\author[G. Malecha]{Gregory Malecha}
\affiliation{
  \institution{BedRock Systems}
  \city{Boston}\state{MA}
  \country{USA}}
\author[B. Pierce]{Benjamin C. Pierce}
\affiliation{
  \institution{University of Pennsylvania}
  \city{Philadelphia}\state{PA}
  \country{USA}}
\author[S. Zdancewic]{Steve Zdancewic}
\affiliation{
  \institution{University of Pennsylvania}
  \city{Philadelphia}\state{PA}
  \country{USA}}

\begin{abstract}
  \textit{Interaction trees} (ITrees) are a general-purpose data
  structure for representing the behaviors of recursive programs that
  interact with their environments.  A coinductive variant of ``free
  monads,'' ITrees are built out of uninterpreted events and their
  continuations. 
  They support compositional construction of interpreters
  from \textit{event handlers}, which give meaning to events by defining
  their semantics as monadic actions.  ITrees are expressive enough to
  represent impure and potentially nonterminating, mutually recursive
  computations, while admitting a rich equational theory
  of equivalence up to weak bisimulation. 
  In contrast
  to other approaches such as relationally specified operational semantics,
  ITrees are executable via code extraction, making them suitable for debugging,
  testing, and implementing software artifacts that are amenable to
  formal verification.

  \bcp{Up to this point, we've made it sound as though ITrees are (a bit)
    novel---a variant on free monads, but not themselves already studied.
    But this sentence sounds like they are ``classic'' structures with known
    theory.  Needs to be clearer.}
  We have implemented ITrees and their associated theory as a Coq library, 
  mechanizing classic domain- and category-theoretic results about program
  semantics, iteration, monadic structures, and equational reasoning.  Although
  the internals of the library rely heavily on coinductive proofs, the
  interface hides these details so that clients can use and reason about ITrees
  without explicit use of Coq's coinduction tactics.

  To showcase the utility of our theory, we prove the termination-sensitive
  correctness of a compiler from a simple imperative source language to an
  assembly-like target whose meanings are given in an ITree-based denotational
  semantics.  Unlike previous results using operational techniques, our
  bisimulation proof follows straightforwardly by structural induction
  and elementary rewriting via an equational theory of combinators for
  control-flow graphs.
\end{abstract}



\maketitle

\section{Introduction}

%


%
Machine-checked proofs are now feasible at scale, for real systems,
in a wide variety of
domains,
including programming language semantics\sz{cite something here?}\bcp{recent
C formalizations? cakeml?} and
compilers~\cite[\etc]{compcert,cakeml}, operating systems~\cite[\etc]{sel4,certikos},
interactive servers~\cite[\etc]{deepweb19}, databases~\cite[\etc]{malecha-db}, and
distributed systems~\cite[\etc]{verdi,ironfleet}, among many
others.\bcp{Mention hardware?  Or is that too big a can of worms?}
%
Common to all of these
is the need to model and reason about interactive, effectful,
and potentially nonterminating computations.
For this, most work to date has relied either on
\textit{operational semantics}, represented as (small- or large-step) transition
relations defined on syntax, or on \textit{trace models},
implemented as predicates over lists or streams of observable events.  These
representations have their advantages: they are expressive, since nearly any
semantic feature can be modeled by transition systems or traces when combined
with appropriate logical predicates; and they fit smoothly with inductive
reasoning
principles that are well supported by interactive theorem provers.  But
they also have significant drawbacks. Operational semantics aren't very
compositional, often requiring auxiliary syntactic constructs (such as
program counters, substitution functions, or evaluation contexts) to specify desired behavior\bcp{not
immediately clear why these are non-compostional};
such syntactic clutter makes proofs unwieldy and brittle.\sz{this phrase is
  trying to succinctly suggest why operational semantics are
  non-compositional.}
Moreover, relational specifications are not executable, which
precludes running the model, either for testing or as a reference
implementation. 

We propose a new alternative called \textit{interaction trees}
(ITrees), a general-purpose data
structure and accompanying theory for
modeling recursive, effectful computations that
can interact with their environment.  ITrees allow us to
give \textit{denotational semantics} for effectful and possibly
nonterminating computations in Gallina, the specification language of Coq~[\citeyear{coq}], despite Gallina's strong purity and termination
constraints.
Such ``shallow''
representations abstract away many syntactic details
and reuse metalanguage features such as function
composition and substitution rather than defining them again, making this
approach inherently more robust to
changes than relational ``deep'' embeddings.
Moreover, ITrees work well with Coq's extraction capabilities,
making it compatible with tools such as QuickChick~\cite{quickchick} for
testing and allowing us to easily link the extracted code against non-Coq
components such as external libraries, so that we can directly execute systems
modeled using ITrees.  This combination of features makes ITrees a good
foundation for formal verification of interactive systems.

The problem of representing effectful programs in pure functional
languages is nearly as old as functional programming itself.  Our design for
interactions trees and their accompanying theory draws heavily on a
large body of prior work, ranging from \textit{monadic
interpreters}~\cite{MoggiMonads89,Steele1994} to 
\textit{free monads}~\cite{Swierstra08} and \textit{algebraic
  effects}~\cite{algebraic-effects}.  Our core data structure, the ITree
datatype itself, is a coinductive variant of the free monad.   Related
structures have been studied previously as the \textit{program
  monad} in the FreeSpec~\cite{freespec} system in Coq, \textit{I/O-trees}~\cite{HS00} and the \textit{general
  monad}~\cite{mcbride-free} in dependent type theory, and the \textit{freer
  monad}~\cite{freer} in Haskell.  \gmm{Perhaps simply: ``ITrees are a
  natural...''. What doesn't seem clear to me is why these two citations should
  be different than the citations in the previous sentence.}\sz{Unlike the
  above, he Delay monad  doesn't support events.  I wanted to mention the category theory
  w.r.t. resumption monads because that is their biggest selling point; it's
  alos pencil and paper (no implementation), unlike the above.}
 ITrees are a natural generalization of Capretta's [\citeyear{Cap05}]
\textit{delay monad} and are an instance of \textit{resumption
  monads}~\cite{coinductive-resumption-monad}, which have been extensively
studied from a category-theoretic point of view.
Section~\ref{sec:related-work} gives a thorough comparison of ITrees with these
and other related approaches.

The use of a coinductive rather than an inductive structure represents a
significant shift in expressiveness, enabling ITrees to represent
nonterminating computations without needing to resort to step-indexed
approaches such as fuel. 
Further, by including a ``silent effect'' (\ilc{Tau}), ITrees can express silently diverging computations and avoid the non-compositionality of guardedness conditions within Coq.
After the fact,\lx{sounds awkward} we can quotient ITrees by these silent steps, providing a generic definition of weak bisimulation.
This also enables us to mechanize classic results from the theory of
iteration~\cite{bloom1993}, which, to our knowledge, have not previously been
applied in the context of machine-checked formalizations.  At a practical level,
this means that we can easily define semantics for mutually recursive
components of an interactive system and reason about them
{compositionally}.

%
Using a coinductive structure comes with some practical tradeoffs.  In
particular, Coq's evaluation of coinductive terms is driven by context, rather
than the term itself, which means that proofs must rely on explicit (or
tactic-driven) rewriting.\gmm{Coq will reduce coinductive definitions when they
  are forced, so I'm not certain that this is entirely accruate. Perhaps we
  should say, ``evaluation of coindutive terms is driven by the context rather
  than by the term itself which requires us to carry out much of our reasoning
  using rewriting.''}\sz{How is this?}
Coq's support for coinductive proofs
is also notoriously limited~\cite{paco}, but we have gone to some pains to
encapsulate the use of coinduction behind the ITrees library interface.  Users of
the ITrees library should rarely, if ever, need to write their own coinductive
proofs.

\paragraph{Contributions} Our main contribution is the design and
implementation of a library that enables formal modeling and reasoning about
interactive, effectful, and potentially nonterminating computations.  Though
it rests on a rich body of existing theory, our work is the first to
simultaneously address four significant challenges. (1) It focuses on
coinductively defined trees whose representation is compatible with program
extraction. (2) It offers a powerful equational theory of monadic
interpreters.
(3) It is realized concretely as a
practical Coq library, paying careful attention to proof-engineering
details that can be
glossed over in pen-and-paper proofs.  And finally, (4) it comes with
a demonstration that the library is usable in practice, in the form of a
novel compiler correctness proof.  Our open-source development is publicly
available,%
\footnote{The link to the ITrees GitHub repository is
  \url{https://github.com/DeepSpec/InteractionTrees}.
  The version of the library as of this publication is on the
  \texttt{popl20} branch.}
 and all of
the results claimed here have been formally proved.  Our experience suggests that ITrees are an effective way to work with
impure and interactive systems in Coq.

The rest of the paper develops these contributions in detail. 

\textit{Section~\ref{sec:itrees}} introduces \textit{interaction trees}
  and establishes that ITrees form a monad with several useful
  notions of equivalence, including variants of strong and weak
  bisimulation.
  It also introduces {\em KTrees} (``continuation ITrees''), a point-free
  representation of functions returning ITrees
  that is convenient for equational reasoning.

\textit{Section~\ref{sec:interpreters}} explains how to
  compositionally give
  semantics to the events of an ITree via monadic \textit{event handlers},
  starting with the familiar example of interpretation into
  the state monad. It then describes the rich algebraic structure of events and handlers
  exposed by the library. \bcp{The two parts of that
    sentence don't obviously have much to do with each other.}
    \lx{I don't understand the second part}

\textit{Section~\ref{sec:recursion}} demonstrates how ITrees support  \textit{recursion and
  iteration}, allowing us to implement a general fixpoint operator,
  \ilc{mrec}, whose properties are also described equationally.
  \sz{Make sure that this outline follows the new structure of the paper.}

\textit{Section~\ref{sec:case-study}} illustrates the use of ITree-based
  denotational semantics with an extended case study.  We \textit{verify the
  correctness of a compiler} from \Imp{} (a simple imperative source language) to
  \Asm{} (a simple assembly language).
  In the example we define the semantics of both languages as ITrees by
  structural recursion on the syntax and the ITree recursion combinators.
  We then prove the equivalence of the denotations by structural induction on the programs, leveraging the ITree library to completely hide the coinductive nature of the proof.
  The final result is a termination-sensitive bisimulation.


\textit{Section~\ref{sec:extraction}} shows (by example) that ITrees are compatible
  with Coq's \textit{extraction} mechanism.  Event handlers can easily be
  written outside of Coq, allowing Coq-generated code to be linked with
  external libraries for the purposes of debugging, testing, and
  implementation.

\textit{Section~\ref{sec:trace}} compares ITrees to more familiar trace-based
  semantics by defining the set of \textit{event traces} of an ITree and
  showing that two ITrees are weakly bisimilar iff their sets of traces
  coincide.  This correspondence means that ITree-based developments can easily
  (and formally) be connected with non-executable models based on small-step
  operational semantics or similar formalisms.

\textit{Section~\ref{sec:related-work}} situates ITrees with respect to
  related work, and
  \textit{Section~\ref{sec:conclusion}} wraps up with a
discussion of limitations and future work.

\section{Interaction Trees}
\label{sec:itrees}

\label{introductiondefinition}

Interaction trees are a datatype for
representing computations that can interact with an external
environment.  We think
of such computations as producing a sequence of \textit{visible
  events}---interactions---each of which might carry a response from the
environment back to the computation.  The computation may also eventually
halt, yielding a final value, or diverge by continuing to
compute internally but never producing a visible event.

\begin{figure}[t]
  \namedsection{positive}
\caption{Simplified presentation of interaction trees.}
\label{fig:itrees_pos}
\end{figure}

Figure~\ref{fig:itrees_pos} shows the definition of the type \ilc{itree E R}.
The parameter \ilc{E : Type -> Type} is a type of \textit{external
  interactions}: it defines the interface by which a computation interacts
with its environment, as we explain below.  \ilc{R} is the \textit{result
  type} of the computation: if the computation ever halts, it will return a
value of type \ilc{R}.\gmm{This entire paragraph essentially talks only about
  the tokens \ilc{itree E R}, which is entirely accurate, the ``intuitive'' part
  comes only in the constructors which are only described in the next
  paragraph. Would it make sense to move the footnote (and possibly the figure
  reference) to the next paragraph?}\sz{Good point.  I moved the footnote
  I think the figure reference here is OK.}

ITrees are defined \textit{coinductively}\footnote{The
    definition shown here follows Coq's historical style of using
    \emph{positive coinductive types}, which emphasizes the tree-like structure
    via its constructors.  This approach is known to break
    subject reduction \cite{Gimenez96} and hence may be deprecated in a future
    Coq version.  Our library therefore uses the recommended \emph{negative
      coinductive} form~\cite{coq-manual:coinduction-caveat,Hagino89} where,
    rather than defining a coinductive type by providing its constructors, we
    instead provide its destructors.  We use ``smart constructors'' for
    \ilct{Ret}, \ilct{Tau}, and \ilct{Vis}, which have the types shown in this
    figure, so the distinction is mostly
    cosmetic (though it does impact the structure of proofs).  We suppress
    these and similar details throughout this paper.} so that they can represent potentially infinite sequences of interactions or divergent behaviors.
They are built using three constructors.
\ilc{Ret r} corresponds to the trivial computation that immediately halts
and produces \ilc{r} as its result. 
\ilc{Tau t} corresponds to a silent step of computation that
does something internal, producing no visible events,
and then continues as \ilc{t}.
Representing silent steps explicitly 
allows ITrees to
represent diverging computations without violating Coq's {\em guardedness
  condition}~\cite{Gimenez05,coinduction}.%
\footnote{%
The guardedness condition is a {syntactic}
side-condition on {\tt cofix} bodies in Gallina. It ensures that a finite
amount of computation suffices to expose the next constructor of the
coinductive type. In practice, it means that the results of co-recursive
calls must occur under constructors and not be eliminated by pattern
matching.\bcp{Can we say that other type-theoretic provers like Agda, Idris,
etc, ... impose similar conditions?  What about Isabelle?}\sz{I believe that
all logically consistent such theorem provers must impose comparable restrictions, but they
may not be exactly equivalent to Coq's.  Agda, at least, has a flag to turn
off termination checking.}\bcp{So shall we add something like ``Other proof
assistants impose related conditions on co-recursive definitions''?}
}

The final and most interesting way to build an ITree is with the
\ilc{Vis A e k} constructor (\ilc{A} is often left implicit).  Here, \ilc{e : E A} is a \emph{visible} external
event, including any outputs provided by the computation to its environment,
and \ilc{A} (for \emph{answer}) is the type of data that the environment provides in response to the event.
The constructor also specifies a continuation,
\ilc{k : A -> itree E T}, which produces the rest of the computation given the
response from the environment.  The tree-like nature of interaction trees stems from the \ilc{Vis} constructor,
since the continuation \ilc{k} can behave differently for different values of
type \ilc{A}.
 Importantly, the continuation is represented as
a meta-level (\textit{i.e.}, Gallina) function, which means both that we can embed
computation in an ITree and that the resulting datatype is extractable and
contains executable functions.

As a concrete example of external interactions, suppose we choose \ilc{E} to
be the following type \ilc{IO}, which represents simple {input/output}
interactions, each carrying a natural number.  Then
we can define an ITree computation \ilc{echo} that loops forever, echoing
each input received to the output:

\vspace{-2ex}
\begin{minipage}[t]{0.35\textwidth}
  \snamedsection{IO}
\end{minipage}
\quad
\begin{minipage}[t]{0.6\textwidth}
  \snamedsection{echo}
\end{minipage}

\noindent Note that \ilc{IO} is indexed by the expected answer type that will be provided by the
environment in each interaction. Conversely, its constructors are
parameterized by the arguments
to be sent to the environment.\gmm{We could probably compress the previous two sentences} Hence, an \ilc{Input} event takes no parameter and
expects a \ilc{nat} in return, while an \ilc{Output} event takes a \ilc{nat} but
expects a non-informative answer, represented by the \ilc{unit} type.
The return type of \ilc{echo} is \ilc{void}, the empty type, since the
computation never terminates.

Similarly, it is easy to define an ITree that silently diverges, producing
no visible outputs and never returning a value:
\snamedsection{spin}
Or one that probes the environment until it receives 9 for an answer, at which
point it terminates (returning \ilc{tt}, the unique value of type \ilc{unit}):
\snamedsection{kill9}

The three basic ITree constructors and explicit \ilc{CoFixpoint} definitions provide very
expressive low-level abstractions, but working with them directly raises several
issues. First, Coq's syntactic guardedness check is inherently
non-compositional, so it is awkward to construct large, complex systems
using it.
Second, we need ways of
composing multiple kinds of events.
Third, 
we often want to model the behavior
of a system by interpreting its events as having \textit{effects} on the
environment.
For example, a \ilc{Write} event could update a memory cell that
a \ilc{Read} event can later access.  Finally, to reason about ITrees and
computations built from them
as above, we would have to use coinduction explicitly.  It is
easier to work with loop and recursion combinators that are more
structured and satisfy convenient equational reasoning principles that can be expressed and proven once and for all.  The
ITrees library provides higher-level abstractions that address all three of
these concerns. 
Figure~\ref{fig:itree-library} provides a synopsis of the library; the details
are explained below.

\vspace{-1ex}
\paragraph{Notation} The library makes extensive use of \textit{parametric functions}, which have
types of the form \ilc{forall (X:Type), E X -> F X}. We write \ilc{E ~>
  F}{} as an
abbreviation for such types.\lx{Put this in figure 2?}


\newcommand{\txt}[1]{\small\textrm{\textit{#1}}}
\newcommand{\stxt}[1]{\footnotesize\textrm{\textit{#1}}}

\begin{figure}
  \begin{minipage}[t]{0.55\textwidth}
\begin{lstlisting}[style=customcoq,basicstyle=\footnotesize\ttfamily]
<@\txt{Interaction tree operations}@>
  itree E A : Type
  Tau  : itree E A -> itree E A
  Ret  : A -> itree E A
  Vis  : {R} (E R) -> (R -> itree E A) -> itree E A
  bind : itree E A -> (A -> itree E B) -> itree E B
  trigger  : E A -> itree E A

<@\txt{Events and subevents}@>
  E, F : Type -> Type
  (e : E R)            R <@\stxt{is the result type of event}@> e
  E +' F               <@\stxt{disjoint union of events}@>
  Typeclass E -< F     E <@\stxt{is a subevent of}@> F
  trigger '{E -< F} : E ~> itree F    <@\stxt{overloaded trigger}@>
\end{lstlisting}
  \end{minipage}
  \begin{minipage}[t]{0.43\textwidth}
\begin{lstlisting}[style=customcoq,basicstyle=\footnotesize\ttfamily]
<@\txt{Heterogeneous weak bisimulation}@>
  eutt (r : A -> B -> Prop) :
  itree E A -> itree E B -> Prop

<@\txt{Strong and weak bisimulation}@>
  _ ~=NUDGE _ : itree E A -> itree E A -> Prop
  _ ~~NUDGE _ := eutt eq.

<@\txt{Parametric functions}@>
  E ~> F  :=  forall (X:Type), E X -> F X

<@\txt{Monadic interpretation}@>
  `{Monad M} `{MonadIter M}
  interp : (E ~> M) -> (itree E ~> M)
\end{lstlisting}
  \end{minipage}

\begin{lstlisting}[style=customcoq,basicstyle=\footnotesize\ttfamily]
<@\txt{Standard event types}@>
  <@\txt{name}@>              <@\txt{events}@>                          <@\txt{handler type}@>
  emptyE               <@\stxt{none}@>                            forall M, emptyE ~> M
  stateE S             Get  Put                        (stateE S) ~> stateT S
  mapDefaultE K V d    Insert  LookupDefault Remove    `{Map K V map} (mapDefaultE K V d) ~> (stateT map)
\end{lstlisting}
\vspace{-1em}

  \caption{Main abstractions of the ITrees library (simplified \&
    abridged).\sz{Should we use \ilc{ktree E A B} instread of \ilc{A -> itree
        E A B} here?} \bcp{Also: Would be really nice to use a larger
      font... (I think this would just require splitting eutt and cat onto
      two lines)}\sz{We may also want to add \ilc{bimap} and
      \ilc{swap} with the heading ``Derived operations''} \bcp{Spacing
      issue: there is more space above this caption than below it!}}
  \label{fig:itree-library}
  \vspace{-1ex}
\end{figure}

\vspace{-1ex}
\subsection{Composing ITree Computations:  ITrees are  Monads}

The type \ilc{itree E} is a {monad}~\cite{MoggiMonads89,monad} for any \ilc{E},
making it convenient to structure effectful computations using the conventions
and notations of pure functional programming.  Figure~\ref{fig:itree-monad}
gives the implementation of the monadic \ilc{bind} and \ilc{ret} operations.  As
shown there, \ilc{bind t k} replaces each~ \ilc{Ret r} with the new
subtree \ilc{k r}.  
%
We wrap the \ilc{Ret} constructor as
a function \ilc{ret} and introduce the usual sequencing notation
\ilc{x <- e ;; k} for bind.

\begin{figure}[t]
\namedsection{monadP}
\begin{lstlisting}[style=customcoq, basicstyle=\footnotesize\ttfamily]
Notation "x <- t1 ;; t2" := (bind t1 (fun x => t2)).
Definition ret x := Ret x.
\end{lstlisting}
  \caption{Monadic \ilc{bind} and \ilc{ret} operators for ITrees.}
  \label{fig:itree-monad}
\end{figure}

We think of the visible events of an ITree as \textit{uninterpreted effects}. In
this sense, \ilc{itree E} is closely related to the {\em free monad} (but
technically distinct: see Section~\ref{sec:related-work}) where every
event of type \ilc{E A} corresponds to an effectful (monadic) operation that can
be ``triggered'' to yield a value of type~\ilc{A}:
\snamedsection{trigger}
Using \ilc{trigger}, we can rewrite the \ilc{echo} example with less syntactic clutter:\gmm{This definition is longer than the original definition so it isn't a very compelling statement.}
%
\snamedsection{echo2}


\subsection{ITree Equivalences}
\label{sec:itree_eq}

Interaction trees admit several useful notions of equivalence even before
we ascribe any semantics to the external events.
These properties are deceptively simple to state, but the weaknesses of coinduction in Coq make some of them quite difficult to prove.

\paragraph{Strong and Weak Bisimulations}

The simplest and finest
notion of equivalence is \textit{strong bisimulation}, written \ilc{t1} $\cong$
\ilc{t2}, which relates ITrees \ilc{t1} and \ilc{t2} when they have exactly the
same shape.

The monad laws and many structural congruences hold up to strong
bisimulation, but once we introduce loops, recursion, or interpreters, which
use \ilc{Tau} to hide internal steps of computation, we need to work with a
coarser equivalence. In particular, we want to equate ITrees that agree on their
terminal behaviors (they return the same values) and on their interactions with
the environment through \ilc{Vis} events, but that might differ in the number of
\ilc{Tau}'s.  This ``{equivalence up to \ilc{Tau}}'' is a form of
\textit{weak bisimulation}: it lets us remove any finite number of
\ilc{Tau}'s when
considering whether two trees are the same, while infinite \ilc{Tau}'s must be
matched on both sides (\textit{i.e.}, this equivalence is termination
sensitive). 
We write \ilc{t} $\approx$ \ilc{u} when \ilc{t} and
\ilc{u} are equivalent up to \ilc{Tau}.  For instance, we have the defining
equation \ilc{Tau t ~~ t}, which does not hold for strong bisimulation but is crucial
for working with general computations modeled as ITrees.

\paragraph{Heterogeneous Bisimulations}

\newcommand{\congR}[1]{\ensuremath{\cong}\raisebox{-.5ex}{#1}}
\newcommand{\approxR}[1]{\ensuremath{\approx}\raisebox{-.5ex}{#1}}

Both strong and weak bisimulation can be further relaxed to relate ITrees that
have different return types, which is needed for building more general
simulations, such as the one used in our compiler correctness proof
(Section~\ref{sec:case-study}).  If we have \ilc{t1 : itree E A} and \ilc{t2 :
  itree E B} and some relation \ilc{r : A -> B -> Prop}, we can define
\ilc{eutt r} (``equivalence up to \ilc{Tau} modulo \ilc{r}''), which is the
same as $\approx$
except that two leaves \ilc{Ret a} and \ilc{Ret b} are related iff
\ilc{r a b} holds.  Intuitively, two such ITrees produce the same external
events and yield results related by \ilc{r}.  Indeed $\approx$ is defined as
\ilc{eutt eq}, where \ilc{r} is instantiated to the Leibniz equality relation \ilc{eq}.  It is
straightforward to generalize $\cong$ in the same way.

Figure~\ref{fig:eq_itree} gives the formal definition of \ilc{eutt r}
as a nested coinductive--inductive structure. The inner inductive \ilc{euttF}
relation is parameterized by \ilc{sim}, a relation on subtrees. It defines a
binary relation on nodes of an ITree demanding that \ilc{Ret a} relates to
\ilc{Ret b} only when \ilc{r a b} holds. Two \ilc{Vis} nodes are related only if
they are labeled with identical events and their continuation subtrees are
related by \ilc{sim} for every value the environment could return, \textit{i.e.}, any
value of type \ilc{R}. \bcp{EqVis should be explained carefully---I'm having
  trouble remembering how to understand it myself right now. (What is u? Why do
  both k1 and k2 take A arguments?)}\yz{There were a couple of typos,
  fixed}\bcp{Still needs an explanation, tho...}
\ilc{EqTau} relates \ilc{Tau t1} and \ilc{Tau t2} whenever \ilc{t1} and \ilc{t2}
are related by \ilc{sim}, while \ilc{EqTauL} and \ilc{EqTauR} allow to strip off
asymmetrically one extra \ilc{Tau} on either side. Note that \ilc{EqTau}
and \ilc{EqVis} appeal \emph{coinductively} to the \ilc{sim} relation whereas
\ilc{EqTauL} and \ilc{EqTauR} appeal \emph{inductively} to \ilc{euttF}. This means
that \ilc{eutt} can peel off only a finite number of \ilc{Tau}'s from one or
both trees before having to align them using \ilc{sim}.
\bcp{Also explain \ilc{<2=}}\gmm{if this is the only place it is used, it is probably better to simply inline it.}

It is easy to show that \ilc{euttF} acts monotonically on relations, which
allows us to define \ilc{eutt} as its greatest fixed point using the \ilc{nu}
operator.  To define \ilc{nu} and to work with coinductive predicates like
\ilc{eutt}, we use the \texttt{paco} library~\cite{paco}, which streamlines
working with coinductive proofs in Coq.

\begin{figure}[t]
  \centering
  \namedsection{eutt1, eutt2, eutt3}
  \caption{Heterogeneous weak bisimulation for ITrees. \bcp{Maybe swap the EqVis
      and EqTau lines, to make the text easier to match up with the figure?}}
  \label{fig:eq_itree}
\end{figure}

Although the definition of heterogeneous weak bisimulation is fairly
straightforward to state, some of its
properties---for instance, transitivity and congruence with respect to
\ilc{bind}---are quite challenging to prove.  For these, we need an appropriate
strengthening of the coinductive hypothesis that lets us reason about \ilc{eutt}
up to closure under transitivity and \ilc{bind} contexts.  Our Coq library
actually uses a
yet more general definition that subsumes both strong and weak bisimulation and
builds in such ``up-to'' reasoning to make proofs smoother;
we omit these details here and refer the interested reader to the Coq
development itself.  The upshot is that we can prove the following:

\ifplentyofspace \begin{theorem}[Equivalences]\ \\ \vspace{-3ex} \else \begin{invisible} \fi
  \begin{enumerate}
  \item $\cong$ is an equivalence relation.
  \item  If \ilc{r} is an equivalence relation, then so is \ilc{eutt r}.
  \item $\approx$ is an equivalence relation (corollary of (2)).
  \item \ilc{t1 ~= t2} implies \ilc{t1 ~~ t2}.
  \end{enumerate}
  \ifplentyofspace \end{theorem} \else \end{invisible}\fi

\paragraph{Equational reasoning}
Fortunately, clients of the ITrees library can treat the definition of \ilc{eutt
  r} and its instances as black boxes---they never need to look at the
coinductive machinery beneath this layer of abstraction.  Instead, clients
should reason equationally about ITrees.  Figure~\ref{fig:eqns}
summarizes the most frequently used equations, each of which corresponds to a
lemma proved in the library.{}   The monad laws, structural laws, and congruences let us
soundly rearrange an ITree computation---typically to put it into a
form where a semantically interesting computation step, such as the
interpretation of an event, takes place\bcp{``takes place'' could be ``comes
  to the outside'' or something?}.  Much of the functionality provided by
the ITrees library involves lifting this kind of equational reasoning to richer
settings, allowing  us to work with combinations of different
kinds of events and interpretations of their effects.

\begin{figure}
  \begin{center}
    \begin{tabular}{lc}
      Monad Laws &
      \begingroup
      \setlength{\tabcolsep}{2pt}
      \renewcommand{\arraystretch}{0.6}
      \begin{tabular}{rcl}
        \ilc{(x <- ret v ;; k x)} & {\small $\cong$} & \ilc{(k v)} \\
        \ilc{(x <- t ;; ret x)} & {\small $\cong$} & \ilc{t} \\
        \ilc{(x <- (y <- s ;; t) ;; u)} & {\small $\cong$} & \ilc{(y <- s ;; x <- t ;; u)}
      \end{tabular}
      \endgroup \\[2em]

      Structural Laws &
      \begingroup
      \setlength{\tabcolsep}{2pt}
      \renewcommand{\arraystretch}{0.6}
      \begin{tabular}{rcl}
        \ilc{(Tau t)} & {\small $\approx$} & \ilc{t} \\
        \ilc{(x <- (Tau t) ;; k)} & {\small $\approx$} & \ilc{Tau (x <- t ;; k)} \\
        \ilc{(x <- (Vis e k1) ;; k2)} & {\small $\approx$} & \\
        \multicolumn{3}{r}{\quad \ilc{(Vis e (fun y => (k1 y) ;; k2))}}
      \end{tabular}
      \endgroup \\[2em]

      Congruences &
      \begingroup
      \setlength{\tabcolsep}{2pt}
      \renewcommand{\arraystretch}{0.6}
      \begin{tabular}{rcrcl}
        \ilc{t1 ~= t2} & {\small $\to$} & \ilc{Tau t1} & {\small $\cong$} & \ilc{Tau t2} \\
        \ilc{k1 ~=~ k2} & {\small $\to$} & \ilc{Vis e k1} & {\small $\approx$} & \ilc{Vis e k2} \\
        \ilc{t1 ~~ t2  NUDGE/\\  k1 ~=~ k2} & {\small $\to$} & \ilc{bind t1 k1} & {\small $\approx$} & \ilc{bind t2 k2}
      \end{tabular}
      \endgroup \\
    \end{tabular}
  \end{center}

  \caption{Core equational theory of ITrees.}
  \label{fig:eqns}
\end{figure}

One pragmatic consideration is that Coq's \ilc{rewrite} and \ilc{setoid_rewrite}
tactics, which let us rewrite using an equivalence (for instance, replacing the
term \ilc{C[t1]} with \ilc{C[t2]} when we know that \ilc{t1} $\approx$
\ilc{t2}), only work if the context is \textit{proper}, meaning that it respects
the equivalence.  Coq's \ilc{Proper} typeclass registers such contexts with the
rewriting tactics. The congruence rules of Figure~\ref{fig:eqns} establish that
the ITree constructors themselves are proper functions.  \bcp{In addition, ?}We prove instances of
\ilc{Proper} for all of the operations, such as those in
Figure~\ref{fig:itree-library}, so that we can rewrite liberally.
Even so, definitions written in monadic style make
heavy use of anonymous functions, which tend to thwart the \ilc{setoid_rewrite}
tactic's ability to find the correct \ilc{Proper} instances.  It is therefore
useful to further raise the level of abstraction to simplify rewriting, as we
show next.

\subsection{KTrees: Continuation Trees}
\label{sec:ktree}

To improve equational reasoning principles and leverage known categorical structures for recursion,
the ITrees library provides an
abstraction for point-free definitions, centered around functions of the form
\ilc{_ -> itree E _}. We can think of these as \textit{impure} Coq functions that
may generate events from \ilc{E} or possibly diverge.
As we will show, they 
enjoy additional structure that we can exploit to generically derive more ways of
composing ITrees computations.
\lx{we could say something about the fact that our categorical abstractions have graphical intuitions (box diagrams as models of monoidal categories)}

We call types of the form \ilc{A -> itree E B} \emph{continuation trees}, or
\emph{KTrees} for short:\lx{This section may be easier to read by dropping
the \ilc{ktree} identifier and using plain arrows \ilc{A -> itree E B} instead.
The main reason for having a definition is for typeclasses.}
\snamedsection{ktreedef}
Whereas an \ilc{itree E R} directly produces an outcome
(\ilc{Ret}, \ilc{Tau}, or \ilc{Vis}), a KTree \ilc{k :} \ilc{ktree E A B} first
expects some input \ilc{a : A} before continuing as an ITree \ilc{(k a)}.
Equivalence on KTrees, written $\eqktree{}$, is defined by lifting weak
bisimulation pointwise to the function space.  \proposecut{With this equivalence
established, we can do proofs by rewriting at the KTree level.}\bcp{Should it
be obvious why this is so, at this point?  Or is this introducing what's
coming next?}

Two KTrees \ilc{h : ktree E A B} and \ilc{k : ktree E B C} can be
composed using \ilc{bind}; the result is written \ilc{(h >>> k) : ktree E A C}.
\snamedsection{ktreecat1,ktreecat2}
KTree composition has a (left and right) identity, \ilc{id_} (equal to \ilc{ret}), and is
associative; the proof follows from the monad laws for \ilc{itree}.
Together, these facts mean that KTrees are the morphisms of a category, the
\emph{Kleisli category}{} of the monad
\ilc{itree E}.

This category has more structure that we expose as part of the ITrees
library interface.  The \ilc{pure} operator lifts a Coq function trivially into
an event-free KTree computation.  We can also easily define an eliminator for the sums type,
\ilc{case_} and corresponding left \ilc{inl_} and right \ilc{inr_}
injections
(effectful variants of the sum type constructors \ilc{inl} and \ilc{inr}).
The names of those operations are suffixed with an underscore so as not to
conflict 
with \ilc{id}, \ilc{inl}, and \ilc{inr} from the standard library,
as well as for the visual uniformity of \ilc{case_} with \ilc{inl_} and \ilc{inr_}.
These operations and their types are summarized in Figure~\ref{fig:itree-library-ktrees}.
They satisfy the equational theory given in Figure~\ref{fig:ktree-eqns}.

\begin{figure}
  \begin{minipage}[b]{0.5\textwidth}
\begin{lstlisting}[style=customcoq,basicstyle=\footnotesize\ttfamily]
id_   : A -> itree E A
cat   : (B -> itree E C) ->
        (A -> itree E B) -> (A -> itree E C)
case_ : (A -> tree E C) ->
        (B -> itree E C) -> (A + B -> itree E C)
inl_  : A -> itree E (A + B)
inr_  : B -> itree E (A + B)
pure  : (A -> B) -> (A -> itree E B)
\end{lstlisting}
  \captionof{figure}{KTree operations}
  \label{fig:itree-library-ktrees}
  \end{minipage}
  \begin{minipage}[b]{0.45\textwidth}
  \setlength{\tabcolsep}{2pt}
  \renewcommand{\arraystretch}{0.6}
  \begin{tabular}{rcl}
    \ilct{id_ >>> k} & {\footnotesize $\eqktree$} & \ilc{k} \\
    \ilct{k >>> id_} & {\footnotesize $\eqktree$} & \ilc{k} \\
    \ilct{(i >>> j) >>> k}  & {\footnotesize $\eqktree$} & \ilc{i >>> (j >>> k)} \\
    \ilct{pure f >>> pure g} & {\footnotesize $\eqktree$} & \ilc{pure (f <o> g)} \\
    \ilct{inl_ >>> case_ h k} & {\footnotesize $\eqktree$} & \ilc{h} \\
    \ilct{inr_ >>> case_ h k}  & {\footnotesize $\eqktree$} &  \ilc{k} \\
    \multicolumn{3}{l}{\ilct{(inl_ >>> f)} {\footnotesize $\eqktree$}
      \ilct{h NUDGE/\\ (inr_ >>> f)} {\footnotesize $\eqktree$} \ilc{k  ->}}\\
    \ilct{f} & {\footnotesize $\eqktree$} & \ilc{case_ h k} \\
  \end{tabular}
    \captionof{figure}{Categorical Laws for KTrees and Handlers
      (\ilc{cat} is denoted by ``\ilc{>>>}'')}
  \label{fig:ktree-eqns}
\end{minipage}
\end{figure}

The laws relating \ilc{case_}, \ilc{inl_}, and \ilc{inr_} mean that KTree is a
\emph{cocartesian category}. The Kleisli and cocartesian categorical structures
are represented using typeclasses.{}  These structures
allow us to derive, generically, other useful
operations and equivalences.  For example,
the following operations \ilc{bimap} and \ilc{swap}
are defined from \ilc{case_}, \ilc{inl_}, and \ilc{inr_}.
The KTree \ilc{bimap f g :} \ilc{ktree E (A + B) (C + D)}
applies the KTree \ilc{f : ktree E A B} if its input is an \ilc{A},
or \ilc{g : ktree E C D} if its input is a \ilc{C};
the KTree \ilc{swap : ktree E (A + B) (B + A)} exchanges the two components
of a sum. 
As we will see below, event handlers also have a cocartesian structure,
which lets us re-use the same generic metatheory for them.

Similarly, the KTree category is just one instance of a Kleisli category, which
can be defined for any monad \ilc{M}.  Monadic event
interpreters, introduced next, build on these structures,
letting us (generically)
lift the equational theory of KTrees to event interpreters too.  This
compositionality is
important for scaling equational reasoning to situations involving many kinds of
events.

\section{Semantics of Events and Monadic Interpreters}
\label{sec:interpreters}

To add semantics to the events of an ITree, we define an \textit{event handler},
of type \ilc{E ~> M} for some monad \ilc{M}.
Intuitively, it defines the meaning of an event of \ilc{E} as a monadic
operation in \ilc{M}. An \textit{interpreter} folds such an event
handler over an ITree; a good interpretation of ITrees is one that respects
\ilc{itree E}'s monadic structure (\textit{i.e.}, it commutes with \ilc{ret} and
\ilc{bind}).

Events and handlers enjoy a rich mathematical structure, a situation well
known from the literature on algebraic effects (see
Section~\ref{sec:related-work}).  Our library exploits this
structure to provide compositional reasoning principles and to lift the base
equational theory of ITrees to their effectful interpretations.

\subsection{Example: Interpreting State Events}
\label{sec:state}

Before delving into the general facilities provided by the ITrees library, it
is useful to see how
things play out in a familiar instance. The code in
Figure~\ref{fig:state} demonstrates how to interpret events into
a state monad.  The event type \ilc{stateE S} defines two events:
\ilc{Get}, which yields an answer of state type \ilc{S}, and \ilc{putE}, which takes
a new state of type \ilc{S} and yields \ilc{unit}.

In the figure, the state monad transformer operations \ilc{getT} and \ilc{putT}
implement the semantics of reading from and writing to the state in terms of the
underlying monad \ilc{M}, using its \ilc{ret}.  The function \ilc{handle_state} is a
\textit{handler} for \ilc{stateE} events: it maps events of type
\ilc{stateE S R}
into monadic computations of type \ilc{stateT S (itree E) R},
\textit{i.e.}, \ilc{S -> itree E (S * R)}, taking an input state to compute an output state
and a result.
Given this handler, we define the \ilc{interp_state} function,
which folds the handler across all of the visible events of an ITree of
type \ilc{itree (stateE S) R} to produce a semantic
function of type \ilc{stateT S (itree E) R}.
The definition of \ilc{interp_state} is an instance of \ilc{interp}
(see Section~\ref{sec:monadicinterpreters} below), specialized to a state monad.
\lx{it should be just \ilc{interp handle_state}, that seems worth inlining}

To prove properties about the resulting interpretation, we
need to show that \ilc{interp_state} is a \textit{monad morphism}, meaning that
it respects the \ilc{ret} and \ilc{bind} operations of the ITree monad.

\ifplentyofspace\begin{lemma}[interp\_state Properties] \hspace{0em} \label{lemma:state}
\else\begin{invisible}\fi
\begin{lstlisting}[style=customcoq, basicstyle=\small\ttfamily]
interp_state (ret x) s  ~~  ret (s, x)
interp_state (x <- t;; k x) s1 ~~ '(s2, x) <- interp_state t s1;; interp_state (k x) s2
\end{lstlisting}
\ifplentyofspace\end{lemma}\else\end{invisible}\fi

We next prove that \ilc{handle_state} implements the desired behaviors for the
\ilc{get} and \ilc{put} operations, which are short-hands for the \ilc{trigger} of the
correponding \ilc{stateE} events.
\begin{invisible}
\begin{lstlisting}[style=customcoq, basicstyle=\small\ttfamily]
interp_state get s ~~ ret (s,s)
interp_state (put s') s ~~ ret (s',tt)
\end{lstlisting}
\end{invisible}
These equations allow us to use \ilc{put} and \ilc{get}'s semantics when
reasoning about stateful computations.
They are also sufficient to derive useful equations when verifying programs
optimizations---for instance, we can remove a redundant \ilc{get} as follows:
\begin{invisible}
\begin{lstlisting}[style=customcoq, basicstyle=\small\ttfamily]
interp_state (x <- get ;; y <- get ;;  k x y) s ~~ interp_state (x <- get ;; k x x) s
\end{lstlisting}
\end{invisible}


\begin{figure}[t]
  \begin{minipage}[t]{0.28\textwidth}
    \namedsection{state1}
  \end{minipage}
  \ \
  \begin{minipage}[t]{0.7\textwidth}
    \namedsection{state1a}
  \end{minipage}
  \begin{minipage}[t]{0.5\textwidth}
  \namedsection{state2}
  \end{minipage}
  \begin{minipage}[t]{0.48\textwidth}
  \namedsection{interpstate}
  \end{minipage}
  \vspace{-1em}
  \caption{Interpreting state events\lx{We could save space by inlining getT and putT}}
  \label{fig:state}
\end{figure}

\subsection{Monadic Interpreters}
\label{sec:monadicinterpreters}

The \ilc{interp_state} function above is an instance of a
general \ilc{interp} function that is defined for any monad \ilc{M},
provided that \ilc{M} supports
an \emph{iteration operator}, \ilc{iter}, of type
\ilc{(A -> M (A + B)) -> A -> M B}.  (The first argument is a loop body
that takes an \ilc{A} and produces either another \ilc{A} to keep
looping with or a final result of type \ilc{B}%
.)
Figure~\ref{fig:interp} shows the definition of \ilc{interp}.
It takes a \ilc{handler : E ~> M} and loops over a tree of type \ilc{itree E
  R}.  At every iteration, the next constructor of the tree is interpreted by
\ilc{handler}, yielding the remaining tree as a new loop state.

The core properties of \ilc{interp}, summarized in Figure~\ref{fig:interp-eqns}, are
generalizations of the laws for \ilc{interp_state}.  In particular, \ilc{interp}
preserves the monadic structure of ITrees, and its action on \ilc{trigger e}
is to apply the handler to the event \ilc{e}.

It remains to show how to instantiate the \ilc{MonadIter} typeclass, which
provides the \ilc{iter} combinator used by \ilc{interp}.  We defer
this discussion to Section~\ref{sec:recursion}, as it will benefit from a
closer look at events and handlers.

\subsection{The Algebra of Events and ITree Event Handlers}
\label{sec:itreehandlers}

The \ilc{handle_state} handler interprets computations with
events drawn from the specific type \ilc{stateE S}.  More generally, we
often want to combine multiple
kinds of events in one computation.  For instance, we might want both
\ilc{stateE S} and \ilc{IO} events, or access to two different types of
state at the same
time. Fortunately, it is straightforward to define \ilc{E +' F}, the
{disjoint union} of the events \ilc{E} and \ilc{F}.  The definition
comes
with inclusion operations \ilc{inl1 : E ~> E +' F} and \ilc{inr1 : F ~> E +'
  F}.\footnote{The \ilct{1} in \ilct{inl1} and \ilct{inr1} reminds us that
  \ilct{E} and \ilct{F} live in \ilct{Type -> Type}.}
 The \ilc{emptyE} event type, with {no} events, is the unit of \ilc{+'}.\gmm{We use \ilc{void} to represent the empty type (in a previous section), but \ilc{emptyE} to represent the empty effect. How hard would it be to be more consistent? And which way should we be more consistent?}

\begin{figure}
  \snamedsection{interp}
  \captionof{figure}{Interpreting events via a handler}
  \label{fig:interp}
\end{figure}

\begin{figure}
  \begin{minipage}[b]{0.53\textwidth}
    \centering
\begin{lstlisting}[style=customcoq,basicstyle=\small\ttfamily]
id_   : E ~> itree E          (* trigger *)
cat   : (F ~> itree G) ->     (* interp *)
        (E ~> itree F) -> (E ~> itree G)
case_ : (E ~> itree G) ->
         (F ~> itree G) -> (E +' F ~> itree G)
inl_  : E ~> itree (E +' F)
inr_  : F ~> itree (E +' F)
\end{lstlisting}
    \captionof{figure}{Event Handler operations}
\label{fig:itree-library-handlers}
\end{minipage}
  \begin{minipage}[b]{0.44\textwidth}
  \begin{center}
  \setlength{\tabcolsep}{2pt}
  \renewcommand{\arraystretch}{1}
  \begin{tabular}{lcl}
    \ilc{interp h (trigger e)} & {\small $\cong$} &\ilc{h _ e} \\
    \ilc{interp h (Ret r)} & {\small $\cong$} & \ilc{ret r} \\
    \ilc{interp h (x <- t;; k x)} & {\small $\cong$} & \\
      \multicolumn{3}{r}{\quad\ilc{x <- (interp h t);; interp h (k x)}} \\
  \end{tabular}
  \end{center}
    \captionsetup{justification=centering}
    \captionof{figure}{
      Some properties of \ilc{interp}.\newline
      See also Figure~\ref{fig:ktree-eqns} for equations in terms of
      \ilc{cat}.}
  \label{fig:interp-eqns}
  \end{minipage}
\end{figure}

The corresponding operations on handlers manipulate sums of event types:
\ilc{case_}
combines handlers for different event types into a handler on their sum, while
\ilc{inl_} and \ilc{inr_} are \ilc{inl1} and \ilc{inr1} turned into event handlers.
\sz{I stated the general rule for handler \ilc{case_}.}

\snamedsection{handler_case}
\vspace{-0.5em}
\begin{lstlisting}[style=customcoq,basicstyle=\small\ttfamily]
Definition inl_ {E F} : E ~> itree (E +' F)
Definition inr_ {E F} : F ~> itree (E +' F)
\end{lstlisting}

Recall that the general type of an event handler is \ilc{E ~> M}. When \ilc{M}
has the form \ilc{itree F}, we can think of
such a handler as \textit{translating} the \ilc{E} events into \ilc{F} events.
We call
handlers of this type \textit{ITree event handlers}.  
%
Like KTrees, event handlers form a cocartesian category where composition of handlers uses
\ilc{interp}, and the identity handler is \ilc{trigger}.
The interface is summarized in Figure~\ref{fig:itree-library-handlers}.

\snamedsection{eventhandler}




\noindent \bcp{Dense!  And this is the first use of \ilc{h ~=~ k}.}
The equivalence relation for handlers \ilc{h ~=~ k} is defined as
\ilc{forall A (e: E A), (h A e) ~~ (g A e)}, \textit{i.e.}, pointwise weak bisimulation.
It admits the same equational theory (and derived constructs) as for KTrees,
hence we reuse the same notations for the operations (see Figure~\ref{fig:ktree-eqns}).

\paragraph{Subevents}
When working with ITrees at scale, it is often necessary to connect ITrees with fewer effects to ITrees with more effects.
For instance, suppose we have an ITree \ilc{t : itree IO A} and we
want to \ilc{bind} it with a continuation \ilc{k} of type \ilc{A -> itree (X +'
  IO +' Y) B} for some event types \ilc{X} and \ilc{Y}.  A priori, this
isn't possible, since the types of their events don't match.  However, since
there is a natural structural inclusion \ilc{inc: IO ~> X +' IO +' Y} (given by
\ilc{inl_ <o> inr_}) we can first interpret \ilc{t} using the handler \ilc{fun e
  => trigger (inc e)} and then bind the result with \ilc{k}.

Since the need for such structural inclusions arises fairly often, the ITrees
library defines a typeclass, written \ilc{E -< F}, that can automatically
synthesize
inclusions such as \ilc{inc}.  It generically derives an instance of \ilc{trigger
  : E ~> itree F} whenever there is a structural subevent inclusion \ilc{E ~>
  F}. We will see in the case study how this flexibility is useful in
practice.

\section{Iteration and Recursion}
\label{sec:recursion}

\begin{figure}
\begin{lstlisting}[style=customcoq,basicstyle=\small\ttfamily]
iter : (A -> itree E (A + B)) -> (A -> itree E B)
loop : (C + A -> itree E (C + B)) -> A -> itree E B
mrec : (E ~> itree (E +' F)) -> (E ~> itree F)
\end{lstlisting}
\caption{Summary of recursion combinators}
\label{fig:itree-library-rec}
\end{figure}

While Coq does provide support for coinduction and corecursion, its technique for establishing soundness relies on syntactic mechanisms that are not compositional.
To make working with ITrees more tractable to clients, our library provides \emph{first-class} abstractions to express corecursion as well as reasoning principles for these abstractions that hide the brittle nature of Coq's coinduction.
From the point of view of a library user, recursive definitions using
these combinators need only to typecheck, even when they lead to divergent behaviors.



Our library exports two iteration constructs, \ilc{iter} and \ilc{loop}, and a
recursion combinator \ilc{mrec} (Figure~\ref{fig:itree-library-rec}).
They are mutually inter-derivable, but they
permit rather distinct styles of recursive definitions.

\subsection{Iteration}

The first function is a combinator for \emph{iteration}, \ilc{iter}, whose implementation is shown in Figure~\ref{fig:iter}.
Given \ilc{body : A -> itree E (A + B)} and a starting state
\ilc{a:A}, \ilc{iter body a} is a computation that
produces either a new state from which to iterate the \ilc{body} again (after a \ilc{Tau}), or a final
value to stop the computation.  This operator makes no assumption on the shape of the loop body, a marked improvement over the intensional guardedness check required by \ilc{cofix}.

\begin{figure}[t]
\begin{minipage}[t]{0.48\textwidth}
  \namedsection{iter}
\end{minipage}
\begin{minipage}[t]{0.50\textwidth}
\namedsection{loopfromiter}
\end{minipage}
  \caption{Iteration combinators: \ilc{iter} and \ilc{loop}.}
  \sz{I suppressed the implicit parameters in these definitions to save
  horizontal space.  We may want to just do that throughout the paper and add a
  mention of that fact somewhere early on.}
\label{fig:iter}
\end{figure}


Defining fixpoint combinators as functions allows
us to prove their general properties \emph{once and for all}.
The equations for \ilc{iter}, given \ifplentyofspace\else below\fi,
imply that continuation trees form an iterative category~\cite{bloom1993}.
The \emph{fixed point identity} unfolds one iteration of the \ilc{iter} loop;
the \emph{parameter identity} equates a loop followed by a computation
with a loop where that computation is part of its last iteration;
the \emph{composition identity} equates a loop whose body sequences
two computations \ilc{f}, \ilc{g} with a loop sequencing them in reverse order,
prefixed by a single iteration of \ilc{f}; and
the \emph{codiagonal identity} merges two nested loops into one.
\ifplentyofspace\begin{lemma}[Iterative category] \label{lemma:iterative} \hfill
\else\begin{invisible}\fi
\vspace{-1em}
  \[\small
    \begin{array}{rclr}
      \ilc{iter f} &  \eqktree & \ilc{f >>> case_ (iter f) id_} & \txt{(fixed point)} \\
      \ilc{iter f >>> g} & \eqktree & \ilc{iter (f >>> bimap id_ g)} & \txt{(parameter)} \\
      \ilc{iter (f >>> case_ g inr_)} & \eqktree & \ilc{f >>> case_ (iter (g >>> case_ f inr_)) id_} & \txt{(composition)} \\
      \ilc{iter (iter f)} & \eqktree & \ilc{iter (f >>> case_ inl_ id_)} & \txt{(codiagonal)} \\
    \end{array}
  \]
\ifplentyofspace\end{lemma}\else\end{invisible}\fi


\sz{It would be good to mention that \emph{proving} these equations is actually  non-trivial and that doing so in the general case in a proof assistant is a significant contribution of this work.  It's also essential, since these kinds of laws are what really buys us expressivity and makes the system useful.}
\lx{Reworded this paragraph}
\noindent The proofs of these equations makes nontrivial use of coinductive reasoning
for weak bisimulation; carrying them out in a proof assistant is a significant contribution
of this work. Nevertheless, that complexity is entirely hidden from users of the library,
behind the simple interface exposed by these equations, whose expressiveness we'll
demonstrate in our case study in Section~\ref{sec:case-study}.

The \ilc{iter} implementation shown in Figure~\ref{fig:iter} is specialized to the \ilc{itree E} monad.  However, we can generalize to other monads and characterize the abstraction using the following typeclass:

\snamedsection{monaditer}
\sz{Should we add this typeclass to Figure~\ref{fig:itree-library}?}

Good implementations of the \ilc{MonadIter} interface must satisfy
\ifplentyofspace
\else the iterative laws. \fi
In a total language such as Coq, this limits the possible implementations.
Base instances include ITrees and the predicate monad (\ilc{_ -> Prop}) where we can tie the knot using the impredicative nature of \ilc{Prop}.
In addition, we can lift \ilc{MonadIter} through a wide variety of monad transfomers, \textit{e.g.}, \ilc{stateT S M} where \ilc{M} is an instance of \ilc{MonadIter}.


\paragraph{Traced categories}
Figure~\ref{fig:iter}\lx{bring this closer} also shows the \ilc{loop}
  combinator, an alternative presentation of recursion that is derivable from
  \ilc{iter}. We can think of the \ilc{C} part of the body's input and output
  types as input and output ``ports'' that get patched together with a
  ``back-edge'' by \ilc{iter}. We use \ilc{loop} in
  Section~\ref{sec:case-study} to model linking of control-flow graphs. This
  \ilc{loop} combinator equips KTrees with the well-studied structure of a
  \emph{traced monoidal category}~\cite{joyal96,Hasegawa97}.

\subsection{Recursion}

Figure~\ref{fig:mrec} shows the code for a general mutual-recursion combinator, \ilc{mrec}.
The combinator uses a technique developed by \citet{mcbride-free} to
represent recursive calls as events. Here, an indexed type
\ilc{D : Type -> Type} gives the signature of a recursive function,
or, using multiple constructors, a block of mutually recursive functions.
For example, \ilc{D := ackermannE} represents a function with two \ilc{nat}
arguments and a result of type \ilc{nat}.
\snamedsection{ackermannE}

A \emph{recursive event handler} for \ilc{D} is an event handler of type
\ilc{D ~> itree (D +' E)}, so it can make recursive calls to itself via
\ilc{D} events, and perform other effects via \ilc{E} events.
As an example, the handler \ilc{h_ackermann} pattern-matches on the event
\ilc{Ackermann m n} to extract the two arguments of the function, and to refine
the result type to \ilc{nat}. The body of the function makes recursive calls by
\ilc{trigger}-ing \ilc{Ackermann} events, without any requirement to ensure the
well-foundedness of the definition.
\snamedsection{ackermannH}

The \ilc{mrec} combinator ties the knot. Given a recursive handler
\ilc{D ~> itree (D +' E)}, it produces a handler \ilc{D ~> itree E}, where all
\ilc{D} events have been handled recursively.
\snamedsection{ackermann}

\begin{figure}[t]
  \centering
  \snamedsection{mreciter}
  \caption{Mutual recursion via events\bcp{The indentation is funny...}}
  \label{fig:mrec}
\end{figure}

The implementation of \ilc{mrec}  in Figure~\ref{fig:mrec}
works similarly to \ilc{interp}, applying the recursive handler \ilc{rh}
to events in \ilc{D}. However, whereas \ilc{interp} directly uses the
ITree produced by the handler as output, \ilc{mrec} adds it as a prefix of the
ITree to be interpreted recursively: the \ilc{inl1 d} branch returns \ilc{bind
  (rh _ d) k}, which will be processed in subsequent steps of
the \ilc{iter} loop.



For reasoning, \ilc{mrec} is also characterized as a fixed point by an
\emph{unfolding equation}, which applies the recursive handler
\ilc{rh : D ~> itree (D +' E)} once,
and interprets the resulting ITree with \ilc{interp}, where \ilc{D}
events are passed to \ilc{mrec} again, and \ilc{E} events are passed to the
identity handler, \textit{i.e.}, \ilc{trigger}, which keeps events uninterpreted.

\begin{invisible}
\vspace{-1em}
  \[\small
\ilc{mrec rh d ~~ interp (case_ (mrec rh) id_) (rh d)}
  \]
\end{invisible}
\lx{TODO: this is sticking out :(}

In fact, \ilc{mrec} is an analogue of \ilc{iter}, equipping event
handlers themselves with the structure of an iterative category. It satisfies the
same equations as \ilc{iter}, relating event handlers instead of KTrees.

\section{Case Study: Verified Compilation of \Imp{} to \Asm{}}
\label{sec:case-study}

To demonstrate the compositionality of ITree-based semantics and the
usability of our Coq library, we use ITrees to formalize and
verify a compiler from a variant of the
\Imp{} language from {\em Software Foundations}~\cite{Pierce:SF1} to a simple
assembly language, called \Asm{}.

We begin by explaining the denotational semantics of \Imp{}
(Section~\ref{sec:imp}) and \Asm{} (Section~\ref{sec:asm}).
It is convenient to define the semantics in stages, each of which
justifies a different notion of program
equivalence.  The first stage maps syntax into ITrees, thereby providing meaning
to the control-flow constructs of the language, but not ascribing any particular
meaning to the events corresponding to interactions with the memory.  The
second stage
interprets those events as effects that manipulate (a
representation of) the actual program state.  

We then give a \emph{purely inductive} proof of the correctness of the
compiler (Sections~\ref{sec:linking}~and~\ref{sec:compiler_correct}).  The
denotational model enables us to state a termination-sensitive bisimulation and
prove it purely equationally in a manner not much different from traditional
compilers for terminating languages with simpler denotational
semantics~\cite{Pierce:SF1}\bcp{funny citation for this}.  The correctness
proof relates the semantics of \Imp{}
to the semantics of \Asm{} \textit{after} their events have been appropriately
interpreted into state monads (a necessity, since compilation introduces new
events that correspond to reading and writing intermediate values\bcp{this
  remark is mysterious at this point}).  Since \Imp{}
programs manipulate one kind of state (global variables) and \Asm{} programs
manipulate two kinds of state (registers and the heap), the proof involves
building an
appropriate simulation relation between \Imp{} states and \Asm{}
states.

To streamline, we identify \Imp{} global
variables with \Asm{} heap
addresses and assume that \Imp{} and \Asm{} programs manipulate the same kinds of
dynamic values.  Neither assumption is critical.

\subsection{A Denotational Semantics for \Imp{}}
\label{sec:imp}

\begin{figure}
\begin{subfigure}[t]{0.45\textwidth}
\begin{lstlisting}[columns=flexible, style=customcoq, basicstyle=\scriptsize\ttfamily]
(* Imp Syntax ------------------------------------ *)
Inductive expr : Set := ... (* omitted *)

Inductive imp : Set :=
| Assign (x : var) (e : expr)
| Seq     (a b : imp)
| If      (i : expr) (t e : imp)
| While   (t : expr) (b : imp)
| Skip.

(* Imp Events ------------------------------------ *)
Variant ImpState : Type -> Type :=
| GetVar (x : var) : ImpState value
| SetVar (x : var) (v : value) : ImpState unit.

Context {E : Type -> Type} `{ImpState -< E}.

(* Imp Denotational semantics -------------------- *)
(* ITree representing an expression *)
Fixpoint denote_expr (e:expr) : itree E value :=
 match e with
 | Var v     => trigger (GetVar v)
 | Lit n     => ret n
 | Plus a b  => l <- denote_expr a ;;
                r <- denote_expr b ;; ret (l + r)
 | ...
 end.
\end{lstlisting}

\end{subfigure}%
~~
\begin{subfigure}[t]{0.55\textwidth}
\begin{lstlisting}[columns=flexible,style=customcoq, basicstyle=\scriptsize\ttfamily,gobble=2]
  (* Imp Denotational semantics cont'd ----------------------- *)
  (* ITree representing an Imp statement *)
  Fixpoint denote_imp (s : imp) : itree E unit :=
    match s with
    | Assign x e   => v <- denote_expr e ;; trigger (SetVar x v)
    | Seq a b      => denote_imp a ;; denote_imp b
    | If i t e     => v <- denote_expr i ;;
        if is_true v then denote_imp t else denote_imp e
    | While t b    =>
        iter (fun _ => v <- denote_expr t ;;
                        if is_true v
                        then denote_imp b ;; ret (inl tt)
                        else ret (inr tt))
    | Skip         => ret tt
    end.

  (* Imp state monad semantics ------------------------------- *)
  (* Translate ImpState events into mapE events *)
  Definition h_imp_state {F: Type -> Type} `{mapE var 0 -< F}
    : ImpState ~> itree F := ...(* omitted *)

  (* Interpret ImpState into (stateT env (itree F)) monad *)
  Definition interp_imp {F A} (t : itree (ImpState +' F) A)
    : stateT env (itree F) A :=
    let t' := interp (bimap h_imp_state id_) t in
      interp_map t'.
\end{lstlisting}
\end{subfigure}
\caption{Syntax and denotational semantics of \Imp{}.  The \ilc{While} case
  uses
  \ilc{iter}; \ilc{GetVar} and \ilc{SetVar} events are interpreted into the
  monad \ilc{stateT env (itree E)}, where \ilc{env} is a finite map from
  \ilc{var} to \ilc{value}.}
\label{fig:imp}
\label{fig:imp-syntax}
\label{fig:imp-semantics}
\end{figure}

The syntax and the semantics for \Imp{} is given in Figure~\ref{fig:imp-syntax}.
In the absence of \texttt{while}, a denotational semantics could be defined, by
structural recursion on statements, as a function from an initial environment to
a final environment; the denotation function would have type \ilc{imp -> env ->
  (env * unit)}.  However, it is not possible to give a semantics to
\texttt{while} using this na\"ive denotation because Gallina's function space is
total.  The usual solution is to revamp the semantics dramatically,
\textit{e.g.}, by moving to a relational, small-step operational semantics
(Section~\ref{sec:compositional-semantics} discusses other
approaches).  With ITrees, the denotation type becomes \ilc{imp -> stateT env
  (itree F) unit}, or, equivalently, \ilc{imp -> env -> itree F (env * unit)},
which allows for nontermination.  It is also more flexible, since the semantics
can be defined generically with respect to an event type parameter \ilc{F}, which
can later be refined if new effects are added to the language or if we
want to compose ITrees generated as denotations of \Imp{} programs with ITrees
obtained in some other way. 

Figure~\ref{fig:imp-semantics} shows how the \Imp{} semantics are structured.  We
first define \ilc{denote_expr} and \ilc{denote_imp}, which result in trees of
type \ilc{itree E unit}. The typeclass constraint \ilc{ImpState -< E}  indicates
that \ilc{E} permits \ilc{ImpState} actions,\bcp{can't parse this sentence---what is a
  refinement of what??} a
refinement of \ilc{stateE} that provides events for reading and writing
\emph{individual} variables;  we would follow the same strategy to add other
events such as IO.  The meanings of expressions and most statements are
straightforward, except for \ilc{While}.  This
relies on the \ilc{iter} combinator (see Section ~\ref{sec:recursion}) to first run the guard expression, then
either continue to loop (by returning
\ilc{inl tt} to the \ilc{iter} combinator) or signal that it is time to stop
(by returning \ilc{inr tt}).

The second stage of the semantics is \ilc{interp_imp}, which takes ITrees
containing \ilc{ImpState} events and produces a computation in the state monad.
It first invokes a handler for \ilc{ImpState}, \ilc{h_imp_state}, to translate the \Imp{}-specific
\ilc{GetVar} and \ilc{SetVar} events into the general-purpose \ilc{mapE} events
provided by the ITrees library (the \ilc{bimap} operator \proposecut{simply ensures
that it} propagates other events untouched). It then uses \ilc{interp_map} to define
their meaning in terms of actual lookup and set operations \proposecut{provided}%
on the type \ilc{env}, a simple finite map from \ilc{var} to \ilc{value}.
%
%
%
%
The final semantics of an \Imp{} statement \ilc{s} is obtained simply by
composing the two functions: \ilc{interp_imp (denote_imp s)}.

Factoring the
semantics this way is useful for proofs.  For instance, to prove the
soundness of a syntactic program transformation from \ilc{s} to \ilc{s'} it
suffices to show that \ilc{denote_imp s ~~ denote_imp s'}; we need not
necessarily consider the impact of \ilc{interp_imp}.
We will exploit this semantic factoring in the
compiler proof below by reasoning about syntactic ``linking'' of \Asm{} code
before its state-transformer semantics is considered.

\bcp{This paragraph could be clearer:}%
This style of denotational semantics avoids defining a syntactic representation of machines, which often comes with a number of administrative reduction rules.
Instead, we represent these administrative reductions using Gallina functions and \ilc{Tau} transitions in the semantics (though these are hidden by \ilc{iter}).
As we will see in Section~\ref{sec:compiler_correct}, this uniform representation will allow us to reason up to \ilc{Tau} and completely ignore these steps in our proofs.

\subsection{A Denotational Semantics for \Asm{}}
\label{sec:asm}

The target of our compiler is \Asm{}, a simple assembly language that
represents computations as collections of basic blocks linked by conditional or
unconditional jumps.
Figure~\ref{fig:asm-syntax} gives the core syntax for the language,
which is split into two levels: basic blocks and control-flow
subgraphs.

\begin{figure}
\begin{subfigure}[t]{0.47\textwidth}
\begin{lstlisting}[columns=flexible,style=customcoq, basicstyle=\scriptsize\ttfamily, gobble=2]
  (* Asm syntax -------------------------------------- *)
  Variant instr : Set := ... (* omitted *)

  Variant branch {label : Type} : Type :=
  | Bjmp (_ : label)                (* jump to label *)
  | Bbrz (_ : reg) (yes no : label) (* cond. jump *)
  | Bhalt.

  Inductive block (label : Type) : Type :=
  | bbi (_ : instr) (_ : block)   (* instruction *)
  | bbb (_ : branch label).        (* final branch *)

  Definition bks A B := fin A -> block (fin B).

  (* Control-flow subgraph: entries A and exits B. *)
  Record asm (A B : nat) : Type :=
  { internal : nat
  ; code      : bks (internal + A) (internal + B) }.

  (* Asm events -------------------------------------- *)
  Variant Reg : Type -> Type :=
  | GetReg (x : reg) : Reg value
  | SetReg (x : reg) (v : value) : Reg unit.

  Inductive Memory : Type -> Type :=
  | Load   (a : addr) : Memory value
  | Store (a : addr) (v : value) : Memory unit.

  Context {E : Type -> Type}.
  Context `{Reg -< E} `{Memory -< E}.

  (* Asm denotational semantics ---------------------- *)
  Definition denote_instr (i:instr) : itree E unit
    := ... (* omitted *)
\end{lstlisting}
\end{subfigure}%
\begin{subfigure}[t]{0.53\textwidth}
\begin{lstlisting}[columns=flexible,style=customcoq, basicstyle=\scriptsize\ttfamily]
(* Asm denotational semantics cont'd ------------------------- *)
Definition denote_br (b:branch (fin B)):itree E (fin B) :=
  match b with
  | Bjmp l => ret l
  | Bbrz v y n =>
      val <- trigger (GetReg v) ;;
      if val ?= 0 then ret y else ret n
  | Bhalt => exit
  end.

Fixpoint denote_bk {L} (b : block L) : itree E L :=
  match b with
  | bbi i b => denote_instr i ;; denote_bk b
  | bbb b    => denote_br b
  end.

Definition denote_bks (bs:bks A B):ktree E (fin A) (fin B)
  := fun a => denote_bk (bs a).

Definition den_asm {A B}:asm A B -> ktree E (fin A) (fin B)
  := fun s => loop (denote_bks (code s)).

(* Asm state monad semantics --------------------------------- *)
Definition h_reg {F: Type -> Type} `{mapE reg 0 -< F}
  : Reg ~> itree F := (* omitted *)
Definition h_mem {F: Type -> Type} `{mapE addr 0 -< F}
  : Memory ~> itree F := (* omitted *)

Definition interp_asm {F A} (t:itree (Reg+'Memory+'F) A)
  : memory -> registers -> itree F (memory*(registers*A)) :=
  let h := bimap h_reg (bimap h_mem id_) in
  let t' := interp h t in
  fun mem regs => interp_map (interp_map t' regs) mem.
\end{lstlisting}
\end{subfigure}
\caption{Syntax and semantics of \Asm{}}
\label{fig:asm}
\label{fig:asm-syntax}
\label{fig:asm-semantics}
\end{figure}

A \emph{basic block} (\ilc{block}) is a sequence of straight-line instructions
followed by a branch that transfers control to another block indicated by a label.
As with \Imp{} expressions, the denotation of instructions is mostly uninteresting, so we omit them.

A \emph{control-flow subgraph}, or ``sub-CFG,'' (\ilc{asm} in
Figure~\ref{fig:asm}) represents the control flow of a computation.
These are \emph{open} program fragments, represented as sets of labeled basic blocks.
The labels in a sub-CFG are separated into three groups:
\emph{entry labels}, from which the code in a sub-CFG can start executing,
\emph{exit labels}, where the control flow leaves the sub-CFG, and
\emph{internal labels}, which are invisible outside of the subgraph.
A sub-CFG has a block of code for every entry and internal
label; control leaves the subgraph by jumping to an exit label. Labels are
drawn from finite domains, \textit{e.g.}\bcp{don't understand ``e.g.'' here:
this is how they are defined.  Also, we should probably explain ``fin'' for
non-Coq experts.}, \ilc{fin A} and \ilc{fin B} where
\ilc{A} and \ilc{B} are \ilc{nat}s, as seen in
\ilc{bks}.\footnote{The finiteness of labels is useful for \Asm{}
  program transformations and is faithful to ``real'' assembly code, but this
  restriction is not actually necessary.  The correctness proof is
  independent of this choice, so we sweep the details under the rug.}
%

\bcp{This paragraph is pretty dense...}%
Figure~\ref{fig:asm-semantics} presents the denotation of \Asm{} programs, which
factors into two parts, just as we saw for \Imp{}.  The \ilc{GetReg} and
\ilc{SetReg} events represent accesses of register state, and \ilc{Load} and
\ilc{Store} accesses of memory.\footnote{ An additional \ilct{Done} event (not
  shown) represents halting the whole program for blocks terminated by a
  \ilct{Bhalt} instruction via \ilct{exit}, but we omit it for the purposes of
  this exposition.}  Once again, we give meaning to the control-flow constructs
of the syntax independently of the state events.  The result of
\ilc{denote_bks} is a \ilc{ktree} that maps each entry label of type \ilc{fin A}
to an \ilc{itree} that returns the label of the next block to jump to.
The \ilc{den_asm} function first computes the denotation of each basic block and
then wires the blocks together using \ilc{loop}, hiding the internal labels in
the process.

The stateful semantics of \Asm{} programs is given by \ilc{interp_asm}, which, like
\ilc{interp_imp}, realizes the register and memory as finite maps using
\ilc{interp_map}.  As a result of this nesting, the ``intermediate
state'' of an \Asm{} computation is a value of type \ilc{memory * (register * A)}.
Because they are built compositionally from interpreters, it is very easy to
prove that both \ilc{interp_imp} and \ilc{interp_asm} are monad morphisms in the
sense that they commute (up to \ilc{Tau}) with \ilc{ret} and \ilc{bind}, a fact
that enables proofs by rewriting.

\subsection{Linking of Control-Flow Subgraphs}
\label{sec:linking}



We now turn to the compilation of \Imp{} to \Asm{}.
The compiler and its proof are each split into two components. The first phase
handles reasoning about control flow by embedding
sub-CFGs into KTrees. In the second phase,
we perform the actual compilation and establish its functional correctness by
reasoning about the quotienting of the local events\bcp{I don't think ``the
  quotienting of the local events'' will mean much to readers at this
  point.}.

For the first phase, we first implement a collection of reusable combinators for
linking sub-CFGs.
These combinators correspond to the operations on KTrees described in
Section~\ref{sec:ktree}\bcp{where should I look for these in
  \ref{sec:ktree}?  I think par\_asm is bimap, but I don't see loop\_asm...?}, which can be seen in this context as presenting a theory of
graph linking at the denotational level.
Here are the signatures of the four essential ones (their implementations
are straightforward):
\begin{lstlisting}[style=customcoq, basicstyle=\small\ttfamily]
Definition app_asm        (ab : asm A B) (cd : asm C D) : asm (A + C) (B + D).
Definition loop_asm       (ab_ : asm (I + A) (I + B))      : asm A B.
Definition pure_asm       (f : A -> B)                     : asm A B.
Definition relabel_asm  NUDGE(f : A -> B) (g : C -> D) (bc : asm B C) : asm A D.
\end{lstlisting}
Two sub-CFGs can be placed beside one another while preserving their labels, via \ilc{app_asm}.
\bcp{The next sentence seems critical, but it's hard to follow. Bit more
  detail please.}\emph{Linking} of compilation units is performed by \ilc{loop_asm}: it connects
a subset of the exit labels \ilc{I} as back edges to the imported labels, also named \ilc{I}, and internalizes them.
Visible labels can be renamed with \ilc{relabel_asm}.
Finally, \ilc{pure_asm} creates, for every label \ilc{a : A}, a block that
jumps immediately to \ilc{f a}.
Together, \ilc{relabel_asm} and \ilc{pure_asm} provide the plumbing required
to use the combinators \ilc{app_asm} and \ilc{loop_asm} effectively.
%

The correspondence between these core \Asm{} combinators and operations on KTrees
is given by the following equations, which commute the denotation function
inside the combinator.
\begin{invisible}
  \vspace{-1em}
\centering
 \[\small
 \begin{array}{rcl}
      \ilc{den_asm (app_asm ab cd)} &  \eqktree & \ilc{bimap (den_asm ab) (den_asm cd)}\\
      \ilc{den_asm (loop_asm ab)} &  \eqktree & \ilc{loop (den_asm ab)}\\
      \ilc{den_asm (relabel_asm f g bc)} &  \eqktree & \ilc{(pure f >>> den_asm bc >>> pure g)}\\
      \ilc{den_asm (pure_asm f)} & \eqktree  & \ilc{pure_ktree f}\\
 \end{array}
 \]
\end{invisible}

\begin{figure}[!t]
\begin{subfigure}[b]{0.6\textwidth}
\centering
\begin{lstlisting}[style=customcoq, basicstyle=\scriptsize\ttfamily]
Definition seq_asm {A B C} (ab : asm A B) (bc : asm B C): asm A C
  :=  loop_asm (relabel_asm swap id_ (app_asm ab bc)).

(* Auxiliary for if_asm *)
Definition cond_asm (e : list instr) : asm 1 (1 + 1)
  := ... (* omitted *)

Definition if_asm {A} (e : list instr) (t : asm 1 A) (f : asm 1 A)
  : asm 1 A
  :=  seq_asm (cond_asm e) (relabel_asm id_ merge (app_asm t f)).

Definition while_asm (e : list instr) (p : asm 1 1) : asm 1 1
  :=  loop_asm (relabel_asm id_ merge
                (app_asm (if_asm e
                           (relabel_asm id inl_ p)
                           (pure_asm inr_))
                         (pure_asm inl_))).
\end{lstlisting}
\end{subfigure}%
~
\begin{subfigure}[b]{0.3\textwidth}
\centering
\begin{tikzpicture}
\node [draw,thick] (ab) at (0,4.4) {ab} ;
\node [draw,thick] (bc) at (0,3.7) {bc} ;
\draw [thick] (ab.east) -- node[midway,above] {$\small{B}$} ($(ab.east) + (0.5,0)$) -- ($(ab.east) + (0.5,0.5)$)
-- ($(ab.west) + (-1.0,0.5)$) -- ($(ab.west) + (-1.0,0)$) -- ($(bc.west) + (-0.3,0)$) -- node[midway,above] {$\small{B}$} (bc.west) ;
\draw [thick] (ab.west) -- ($(ab.west) + (-0.3,0)$) -- ($(bc.west)  + (-1.0,0)$) -- node[midway,above] {$\small{A}$} ($(bc.west)  + (-1.5,0)$) ;
\draw [thick] (bc.east) -- node[midway,above] {$\small{C}$} ($(bc.east) + (1,0)$);
%
\node [draw,thick] (tp) at (0.8,2.9) {t} ;
\node [draw,thick] (fp) at (0.8,2.1) {f} ;
\node [draw,thick] (ife) at (-1,2.5) {e} ;
\draw [thick] ($(ife.west) + (-1,0)$) -- node[midway,above] {1} (ife.west);
\draw [thick] (ife.east) -- node[midway,above] {true}  (tp.west) ;
\draw [thick] (ife.east) -- node[midway,below] {false} (fp.west) ;
\draw [thick] (tp.east) -- node[midway,above] {A} ($(tp.east) + (0.5,0)$) -- ($(tp.east)+(0.8,-0.4)$) -- node[midway,above] {A} ($(tp.east)+(1.3,-0.4)$);
\draw [thick] (fp.east) -- node[midway,below] {A} ($(fp.east) + (0.5,0)$) -- ($(fp.east)+(0.8,+0.4)$);
%
\node [draw,thick] (p) at (0.8,0.9) {p} ;
\node [draw,thick] (e) at (-1,0.5) {e} ;
\draw [thick] (p.east) -| ++(0.4,0.5) -| ($(e.west)-(0.4,0)$) -- (e.west) ;
\draw [thick] (e.east) -- node[midway,above] {true} (p.west) ;
\draw [thick] (e.east) -- node[midway,below] {false} ($(p.west)-(0,0.8)$) -- node[midway,above] {1} ++(1.5,0) ;
\draw [thick] (e.west) -- node[midway,below] {1} ($(e.west) - (1,0)$) ;
\end{tikzpicture}
\end{subfigure}
\caption{High-level control flow in \Asm{}\bcp{Let's make the drawings a bit
    smaller.  And there are blank lines on the left that can be tightened
    once the drawing is small.}}
\label{fig:asm-high-combinators}
\end{figure}

Equipped with these primitives, building more complex control-flow graphs
becomes a \proposecut{very} diagrammatic game.
Figure \ref{fig:asm-high-combinators} shows how to use the primitives to build
linking operations for sub-CFGs that mimic the control-flow operations provided by \Imp{}.
For instance, sequential composition of \ilc{asm A B} with \ilc{asm B C}
places them in parallel, swaps their entry labels to get a
sub-CFG of type \ilc{asm (B+A) (B+C)}, and then internalizes the intermediate label \ilc{B} via \ilc{loop_asm}.

\bcp{Confusing: seems like something can't be both specific and generic.}We emphasize that while these control-flow graphs are
specific to \Imp{}, their definitions do not depend on
\Imp{}'s or even \Asm{}'s state-transformer semantics.  We can reason about
control-flow independently of other events.
For instance, the denotation of the \ilc{seq_asm} combinator
is indeed the sequential composition of denotations of its arguments (up to
\ilc{Tau}):
\begin{lstlisting}[style=customcoq, basicstyle=\small\ttfamily]
Lemma seq_asm_correct {A B C} (ab : asm A B) (bc : asm B C) :
  (den_asm (seq_asm ab cd)) ~=~ (den_asm ab >>>  den_asm bc).
\end{lstlisting}

The \ilc{while_asm} combinator is, naturally, more involved.
As illustrated in Figure~\ref{fig:asm-high-combinators},
it constructs the control-flow graph of a \texttt{while} loop given
the list of instructions for the test condition and the compilation
unit corresponding to the body of the loop.
The type of \ilc{p} represents a compilation unit with a single imported
label (the target to jump to when the loop body finishes) and a single exported label (the entry label for the top of the loop body).
The correctness of the combinator establishes that its denotation can be viewed
as an entry point that runs the body if a variable \ilc{tmp_if} is non-zero
after evaluating the expression \ilc{e}.
This is expressed at the level of KTrees via the \ilc{loop} operator.
In the code below, \ilc{label_case l} analyzes the
shape of the label \ilc{l}, and \ilc{l1} and \ilc{l2} are two distinct label
constants corresponding to the loop entry or exit, respectively.
\begin{lstlisting}[style=customcoq, basicstyle=\small\ttfamily]
Lemma while_asm_correct (e : list instr) (p : asm 1 1)
  : denote_asm (while_asm e p)
  ~=~ loop (fun l:fin (1 + 1) =>
       match label_case l with
       | inl _ =>  denote_list e ;; v <- trigger (GetReg tmp_if) ;;
                   if (v:value) then Ret l2 else (denote_asm p l1;; Ret l1)
       | inr _ => Ret l1
       end).
\end{lstlisting}
Most importantly, the proof of \ilc{while_asm_correct} is again purely equational,
relying solely on the theory of KTrees and correctness equations of the low-level linking
combinators (\ilc{app_asm_correct}, \ilc{seq_asm_correct}, \etc).\sz{This is a place where we could mention \ilc{eutt_clo_bind} if we
  wanted to, it's very useful in this proof.}


\subsection{Compiler Correctness}
\label{sec:compiler_correct}

The compiler itself is entirely
straightforward.  It compiles \Imp{} statements using the
linking combinators along with \ilc{compile_assign} and \ilc{compile_expr},
both of which are simple (and omitted). We pass to \ilc{compile_expr}
the name of a target register (here, just \ilc{0}), into which the value of the
expression will be computed; it stores intermediate results in additional \Asm{}
registers as needed.  \bcp{Should be able to save a line or two by
  compressing the code here:}

\pagebreak[3]
\begin{lstlisting}[style=customcoq, basicstyle=\small\ttfamily]
Fixpoint compile (s : stmt) {struct s} : asm 1 1 :=
  match s with
  | Skip         => id_asm
  | Assign x e   => raw_asm_block (after (compile_assign x e) (Bjmp l1))
  | Seq l r      => seq_asm (compile l) (compile r)
  | If e l r     => if_asm (compile_expr 0 e) (compile l) (compile r)
  | While e b    => while_asm (compile_expr 0 e) (compile b)
  end.
\end{lstlisting}


The top-level compiler correctness theorem is
phrased as a {bisimulation} between the \Imp{} program and the corresponding
\Asm{} program, which simply requires them to have ``equivalent'' behavior.
\begin{lstlisting}[style=customcoq, basicstyle=\small\ttfamily]
Theorem compile_correct (s : stmt) :  equivalent s (compile s).
\end{lstlisting}
\bcp{Dense:}Figure~\ref{fig:correctness-prerequisites} unpacks the definition of
\ilc{equivalent}, which requires the ITree denotations of \ilc{s} and its
compilation to be bisimilar. Two ITrees \ilc{t1} and \ilc{t2},
representing \Imp{} and \Asm{} computations respectively,
of types \ilc{itree (ImpState +' E) A} and \ilc{itree (Reg +' Memory +' E) B},
are bisimilar if, when run in
\ilc{Renv}-related initial states, they produce computations that are equivalent
up to \ilc{Tau} and both terminate in states related by \ilc{state_invariant TT}.
The relation \ilc{Renv} formalizes the assumption that the \Imp{}
environment and \Asm{} memory have the same contents when viewed as maps from \Imp{}
variables / \Asm{} addresses to values, and it is implied by
\ilc{state_invariant}.
Here, \ilc{TT} is the trivial relation on
the output label of \Asm{}, since a statement has a unique exit point; in general the
\ilc{RAB} relation parameter in \ilc{state_invariant} is used to ensure that both
computations jump to the same label, which is needed to prove that loops
preserve the state invariant.

Since the compiler introduces temporary variables, the bisimulation does not
hold over the uninterpreted ITrees.\sz{I guess that we might be able to get away
  with interpreting just the \ilc{Reg} events on the \Asm{} side if we translate
  \ilc{ImpState} and \ilc{Memory} to \ilc{MapDefaultE}.  It would be
  interesting to do this experiment to see how much simpler the proofs become.}  To prove that expressions are compiled
correctly, we need to explain how reads and writes of \Asm{} registers relate to
the computations done at the \Imp{} level.  The relation \ilc{sim_rel} establishes
the needed invariants, which ensure that the code generated by the compiler
(1) doesn't corrupt the \Asm{} memory, (2) uses registers in a ``stack
discipline,'' and (3) computes the \Imp{} intermediate result \ilc{v} into the
target register \ilc{n}.  These properties are used to prove the correctness of
\ilc{compile_expr}.




\begin{figure}
\begin{lstlisting}[columns=flexible, style=customcoq, basicstyle=\footnotesize\ttfamily]
(* Relate an Imp env to an Asm memory *)
Definition Renv (g_imp : Imp.env) (g_asm : Asm.memory) : Prop :=
  forall k v, alist_In k g_imp v IFF alist_In k g_asm v.

Definition sim_rel l_asm n: (Imp.env * value) -> (Asm.memory * (Asm.registers * unit)) -> Prop :=
    fun '(g_imp', v) '(g_asm', (l_asm', _))  =>
      Renv g_imp' g_asm' /\            (* we don't corrupt any of the Imp state *)
      alist_In n l_asm' v /\           (* we get the right value in register n *)
      (forall m, m < n -> forall v,   alist_In m l_asm v IFF alist_In m l_asm' v).
                                      (* we don't mess with anything on the "stack" *)

Context {A B : Type}.              (* Imp / Asm intermediate result types *)
Context (RAB : A -> B -> Prop).  (* Parameter that relates intermediate results *)

(* Relate Imp to Asm intermediate states. *)
Definition state_invariant (a : Imp.env * A) (b : Asm.memory * (Asm.registers * B))  :=
    Renv (fst a) (fst b) /\ (RAB (snd a) (snd (snd b))).

Definition bisimilar {E} (t1 : itree (ImpState +' E) A) (t2 : itree (Reg +' Memory +' E) B)  :=
    forall g_asm g_imp l, Renv g_imp g_asm
    -> eutt (state_invariant RAB) (interp_imp t1 g_imp) (interp_asm t2 g_asm l).

(* Imp / Asm program equivalence *)
Definition TT : unit -> fin 1 -> Prop := fun _ _ => True.
Definition equivalent (s:stmt) (t:asm 1 1) : Prop := bisimilar TT (denote_stmt s) (den_asm t f1).
\end{lstlisting}
  \caption{The simulation relations for the compiler correctness
    proof. \sz{Annoyingly, we've defined \ilc{sim_rel} in a way that is syntactically
      different from \ilc{state_invariant}} \bcp{Save a line by combining
      the two Context declarations.  Maybe save a line by not putting third ``we
      don't mess...'' on the line it refers to (and shortening it)}}
\label{fig:correctness-prerequisites}
\end{figure}


Crucially, despite correctness being termination sensitive,
the proofs follow by structural induction on the \Imp{} terms: all coinductive
reasoning is hidden in the library. As in the first phase, the reasoning here
follows by rewriting, this time using the
\ilc{bisimilarity} relation and equations about \ilc{interp_imp} and
\ilc{interp_asm} that are induced by virtue of being compositionally
defined from \ilc{interp_state}.

Setting aside the usual design of the simulation relation, the resulting proofs
are slightly verbose, but extremely elementary. They mostly
consist in successive rewrites to commute the denotation with the various
combinators, and some elementary semantic reasoning where events are reached to
prove that the simulation relation is preserved.
We believe that these kind of equational proofs can be automated to a large
degree, a perspective we would like to explore in further works.

\proposecut{The proofs rely on several auxiliary lemmas,
but, in each case, the correctness of the linking combinator and the equational
theory of KTrees permit us to reduce the goal to reasoning about the basic
interpretations of the events.}

\section{Extracting ITrees}
\label{sec:extraction}

One of the big benefits of ITrees is that they work well with Coq's extraction
facilities.  If we extract the \ilc{echo} definition from Section~\ref{sec:itrees},
we obtain the code shown at the top of Figure~\ref{fig:extract}.\footnote{For
  simplicity, here we also extract Coq's \ilc{nat} type as OCaml's \ilc{int}
  type.} The \ilc{itree} type extracts as a lazy datatype and \ilc{observe}
forces its evaluation.


To actually run the represented computation, we provide a driver that
traverses the \ilc{itree}, forcing all of its computation and providing handlers
for any visible events that remain in the tree.  The OCaml function \ilc{run}
does exactly that, where, for the sake of this example, we interpret each
\ilc{Input} event as a call to OCaml's \ilc{read_int} command and each
\ilc{Output} event as a call to \ilc{print_int}.\footnote{
  Thanks to its dependent type, the OCaml extraction of \ilct{Vis} uses OCaml's
  \ilct{Obj.t} as the domain of the embedded continuations, so handlers should be
  written with care, 
  otherwise type-safety could be jeopardized.} This kind of simple event
handling already suffices to add basic IO and ``printf debugging'' to Coq programs, which can
be extremely handy in practice.  We can, of course, implement
more sophisticated event handlers, using the full power of OCaml.

\begin{figure}[t]
  \centering
\begin{lstlisting}[style=customocaml, basicstyle=\small\ttfamily,linerange={16-18}]
  let Go observe0 = Lazy.force i in observe0

type 'x iO =
\end{lstlisting}
\begin{lstlisting}[style=customocaml, basicstyle=\small\ttfamily,linerange={12-21}]
let handle_input k = k (Obj.magic (read_int ()))
let handle_output x k = print_int x ; k (Obj.magic ())

(* Driver *)
let rec run t =
  match observe t with
  | RetF r -> r
  | TauF t -> run t
  | VisF (Input, k) -> handle_input (fun x -> run (k x))
  | VisF ((Output x), k) -> handle_output x (fun x -> run (k x))
\end{lstlisting}
  \caption{OCaml extracted from the \ilc{echo} example (top); OCaml
    handler and ``driver'' loop (bottom).}
  \label{fig:extract}
\end{figure}


ITrees extractability has played a key role in several different parts of
an ongoing research project that seeks to use Coq for \textit{Deep
Specifications}.\footnote{\url{http://www.deepspec.org}.}
In particular,  our re-implementation of the Vellvm\footnote{\url{https://github.com/vellvm/vellvm}} formalization of LLVM,
which aims to give a formal semantics for the LLVM IR in Coq, heavily
uses ITrees exactly as proposed in this paper to build a denotational semantics
for LLVM IR code.  The Vellvm semantics has
many layers of events and handlers (for global data, local data, interactions
with the memory model, internal and external functions calls, \etc), and the
LLVM IR control-flow graphs are a richer version of the \Asm{} language
(Section~\ref{sec:asm}).  The flexibility of using ITree-based interpreters means that
Vellvm can define a relational specification that accounts for nondeterministic
features of the LLVM (such as \texttt{undef}) but that can also be refined into
an implementation.
We are able to extract an executable interpreter that performs
well enough to test small- to medium-sized LLVM code samples (including
recursion, loops, \etc).  All but the outermost \ilc{run} driver are extracted
from Coq, as in the \ilc{echo} example.

We are also using ITrees as executable specifications to model the semantics of
web servers \cite{deepweb19}.  The ITree representation serves two purposes:
(1) ITrees model the interactive operations of the web server in a way that can
be connected via Princeton's VST framework \cite{vst,vst-paper} to a C
implementation, and (2) the model can also be used for property-based testing
with QuickChick~\cite{quickchick}. The ability to link against
handlers written in OCaml means that the testing framework can be used to test
real web servers like Apache across the network, in addition to linking against
our own web servers.  Here again, the performance of the extracted executable
has been good enough that we have felt no need to do any optimization on the
ITree representation.


\section{Relating ITrees and Trace Semantics}
\label{sec:trace}

We have shown that ITrees provide a way to define denotational
semantics for possibly diverging, effectful programs in Coq.  In this approach,
we use monadic interpreters that produce ITrees as a semantic
representation of program behaviors, which we can then reason
about equationally.

A more common approach to defining language semantics in Coq (and in other proof
assistants) is via a \emph{deep embedding}, in which the program's operational
semantics are specified relationally.  For
example, the CompCert project~\cite{compcert}  takes this approach, where
the fundamental transition relation is given
by \ilc{step : state -> event -> state -> Prop}.
Here, a \ilc{state} is a representation of the current program state,
and the
\ilc{event} type contains information about both the outputs to the environment
and the inputs that the program might receive from the environment.  The
proposition \ilc{step s1 e s2} holds when it is possible for the system to
transition from state \ilc{s1} to state \ilc{s2} while producing the observable
event \ilc{e}.  The meaning of a complete program is given by the set of all
finite sequences of events, called \textit{traces}, generated by
the transitive closure of the step relation.

An important distinction in trace semantics is that the \ilc{step}
relation \textit{quantifies} over possible inputs
that it might receive from the environment.  Unlike ITrees, whose \ilc{Vis e k}
constructors expose the continuation \ilc{k} as a \textit{function}, relational
semantics with input events are fundamentally not executable, because of that
universal quantification.  If we try to shoehorn ITrees into a small-step
transition relation in this style, the \ilc{Vis} case becomes:

\snamedsection{step_vis}

We instantiate the \ilc{event} type as an existential package containing a pair of
the event \ilc{e : E X} and a response \ilc{x : X}, which is universally
quantified---the step can take place for any \ilc{x} of
type \ilc{X} provided by the environment.  This propositional encoding of
inputs means that operational
semantics developed in this style cannot be extracted from Coq\bcp{why?}.  Consequently,
such semantic definitions cannot easily be used as implementations or
executable tests.
The CompCert project goes to some pains to implement a separate interpreter
that corresponds to their small-step semantics to aid with debugging.%

On the other hand, traces are sometimes convenient, particularly when there is
inherent nondeterminism in the specification of a system's behaviors.%
\bcp{why
  particularly then? is there something wrong with the ITree presentation of
  nondeterminism?}\sz{I was thinking that nondeterminism in trace models doesn't
  usually count choice as an observable event.  That means that the natural
  trace models of $a \oplus b$ and $(a \oplus b) \oplus a$ should be equivalent
  since they have the same set of visible traces, namely $\{a,b\}$.  If choice
  is a visible event $?$, then we get the set of traces $\{?a, ?b\}$ for the
  first one and $\{??a, ??b, ?a\}$ for the second, which isn't the same---we
  need to quotient by treating $?$ just like \ilc{Tau}.  That said, we could
  still reason about nondeterminism in ITrees, but we might want to do it up to
  refinement, not equivalence.}  Rather than defining a trace of an
ITree using a small-step semantics, like CompCert, it is more natural to think
of an ITree as directly denoting a set of possible traces---finite prefixes of
paths through the tree that record \ilc{Vis} events and corresponding responses
from the environment.  This definition is shown in Figure~\ref{fig:trace}.

\begin{figure}[t]
\center
\namedsection{trace}
\caption{ITree traces: \ilc{is_trace_of t tr} means that \ilc{tr} is a possible
  trace of ITree \ilc{t}. }
\label{fig:trace}
\end{figure}

The \ilc{trace} datatype is intuitively a list of events.  \ilc{TEnd} marks a
partial trace; it corresponds to \ilc{spin} or $\bot$ (we cannot distinguish the
two, a problem of using finite traces). \ilc{TRet r} denotes a computation that
finished, producing the value \ilc{r}.  \ilc{TEventResponse e x t} corresponds
to a \ilc{Vis} event \ilc{e} to which the environment responded with the answer
\ilc{x}, then continues with trace \ilc{t}, and \ilc{TEventEnd e} corresponds
to a situation in which the ITree is waiting for a response from the environment
(perhaps one that will never come, \textit{i.e.}, if it has an event of type
\ilc{E void}).

The \ilc{is_trace_of} predicate leads to a natural notion of trace refinement, and thus
a different characterization of ITree equivalence.


\begin{definition}[Trace Refinement]
  \ilc{t} $\sqsubseteq$ \ilc{u} iff \ilc{forall tr, is_trace_of t tr -> is_trace_of u tr}.
\end{definition}

\begin{definition}[Trace Equivalence]
  \ilc{t} $\equiv$ \ilc{u} iff \ilc{t} $\sqsubseteq$ \ilc{u} and \ilc{u} $\sqsubseteq$ \ilc{t}.
\end{definition}

Using these definitions, we can show that trace equivalence coincides with
weak bisimulation\ifplentyofspace
\else, \textit{i.e.}, that \ilc{t1} $\approx$ \ilc{t2} $\iff$ \ilc{t1} $\equiv$ \ilc{t2}.
\fi

\bcp{I don't understand much in the following paragraph:}%
Trace refinement shows that \ilc{spin} $\sqsubseteq$ \ilc{t}.  Since all events
in the ITree are visible in the trace, if \ilc{u} $\sqsubseteq$ \ilc{t}, then
\ilc{t} can be obtained from \ilc{u} by replacing silently diverging behaviors
with some visible events, or otherwise inserting/removing a finite number of
\ilc{Tau}s into/from \ilc{u}. In many cases, different notions of refinement may
be more useful. For instance, one may want nondeterministic behavior to be
refined by possible deterministic behaviors. One way to do this is to directly
allow the \ilc{is_trace_of} predicate to allow a choice of possible refinements,
effectively interpreting away some events in the refinement definition.  A
second option is to introduce a more sophisticated version of refinement that
uses both the definition above and an event interpreter as described in
Section~\ref{sec:monadicinterpreters}---the resulting relations should be quite similar
to those studied by \citet{JSV10}.

\gmm{the distinction between traces and trees is reminiscent of the distinctions between branching and linear time logics which are, if i recall correctly, incomparable in expressivity. this might not be something that we can get into this late in the game though.}
\bcp{Are you saying that there are two programs that can be distinguished by
  trace equivalence but not by (strong or weak) bisimulation?  That would
  surprise me---got an example?}\sz{Bisimulation is, in general, finer than
  trace equivalence, I thought. Google turned up this paper ``Deciding
  bisimulation and trace equivalences for systems with many identical
  processes'' that might shed light on your question.}





\section{Related Work}
\label{sec:related-work}

The problem of accommodating effectful programming in purely functional
settings is an old one, and a variety of approaches have been explored,
monads and algebraic effects being two of the most prominent.  We
concentrate on these two techniques, beginning with general background and
then focusing on the closest related work.

\bcp{Writing nit: We are inconsistent about grammatical tense when
  we talk about other work, switching back and forth between past, past
  participle, and present.  All of these are OK (well, past participle is a
  bit clunkier than the others), but we should be consistent.  (Actually,
  now that I think of it, past tense is better, because then if we switch
  to present tense it is clear that we are talking about our own work.)}

\subsection{Monads, Monad Transformers, and Free Monads}

Moggi's seminal paper~[\citeyear{MoggiMonads89}] introduced monads as one way to give
meaning to imperative features in purely functional programs. Monads were
subsequently popularized by \citet{monad} and \citet{PJW93} and have had
huge impact, especially in Haskell.  However, it
was soon recognized that composing monads to combine multiple
effects was not straightforward.  Monad transformers \cite{Moggi90} are
one way to obtain more compositionality; for example, \citet{LHJ95} showed
how they can be used to build interpreters in a modular way.
The \ilc{interp_state} function from Section~\ref{sec:interpreters}
is an example of building an event interpreter using a monad transformer in this
style. In our case, not all monads are suitable targets for interpretation:
we require them to support recursion in the sense that their Kleisli category
is iterative.
Correspondingly, not all monad transformers can therefore be used to build
interpreters.

Sweirstra's  Datatypes \`{a} la Carte [\citeyear{Swierstra08}] showed
how to use a \textit{free monad} to define monad instances modularly.
Transporting his definition to our setting, we would obtain the following:

\begin{lstlisting}[style=customcoq, basicstyle=\small\ttfamily]
CoInductive Free (E : Type -> Type) (R:Type) :=
| Ret : R -> Free E R
| Vis : E (Free E R) -> Free E R.
\end{lstlisting}

This version of the \ilc{Vis} constructor directly applies the functor
\ilc{E} to the coinductively defined type \ilc{Free E R} itself.
However, this type violates the strict positivity condition enforced by Coq:
certain choices of \ilc{E} would allow one to construct an infinite loop.
\proposecut{(even if we replace \ilc{CoInductive} by \ilc{Inductive} above)}\bcp{Is there a citation for that observation?  Or could one say
  just a few more words about which choices?}\yz{I think "Structural Recursive
  Definitions in Type Theory" by Gimenez could be adequate. Alternatively, his
  PhD}

Subsequent work by \citet{apfelmus}, \citet{KSS13} and \citet{freer} showed how free monads can
be made more liberal by exposing the continuation in the \ilc{Vis} constructor.
The resulting ``freer'' monad (called \ilc{FFree} in their work) is
essentially identical to
our ITrees---the difference being that, because they work in Haskell,
which admits
nontermination by default, it needs no \ilc{Tau} constructor.

When considered up to strong bisimulation, ITrees form the free \emph{completely
iterative monad}~\cite{aczel2003} with respect to a functor of the form
\inlinecoq{fun X => F X + X}, where the second component corresponds to
\inlinecoq{Tau} nodes.
Quotiented by weak bisimulation, ITrees define a free \emph{pointed
monad}~\cite{Uustalu17}.
There is a rich literature on the theory of iteration
\cite{bloom1993,milius2005,goncharov2017}, studying the properties of operators
such as \inlinecoq{mrec} in yet more general category-theoretic settings.  The
ITrees library makes such results concretely applicable to formally verified systems.

ITrees are a form of resumptions, which originated
from concurrency theory~\cite{milner1975}. More precisely, ITrees can
be obtained by applying a coinductive resumption monad
transformer~\cite{coinductive-resumption-monad,cenciarelli1993} to the delay monad of
\citet{Cap05}.\sz{Is that true?  How does the resumption monad transformer
  inject \ilc{Vis} nodes into the coinductively defined delay type?}\lx{The resumption monad transformer alternates layers of "external effects" (E events) and "internal effects" (divergence in the Delay monad) \ilc{itree F := nu T. Delay(F T + R)}}
  Other variations of the resumption monad transformer have been
used to model effectful and concurrent
programs~\cite{nakata2010,goncharov2011}.
In particular, \citet{nakata2010} also used coinductive resumptions in Coq to
define the semantics of \Imp{} augmented with input-output operations. They also
defined termination-sensitive weak bisimilarity (``equivalence up to taus'') using
mixed induction-coinduction.  However, their semantics was defined as an
explicitly coinductive \textit{relation}, with judicious introductions of
\ilc{Tau}. Their Coq development was specialized
to \Imp{}'s global state and was not intended to be used as a general-purpose
library.  In contrast, our semantics are functional (denotational, definitional)
\textit{interpreters}, and we encapsulate nontermination (\ilc{Tau} is an internal
implementation detail) using recursion
operators that are compatible with Coq's extraction mechanisms.
\lx{This seems a bit long winded}
%
%


\subsection{Algebraic Effects and Handlers}

Algebraic effects are a formalism for expressing the semantics of effectful
computations based on the insight by Plotkin and Power that many
computational effects are naturally described by algebraic theories
[\citeyear{PP01,PP02,algebraic-effects}].  The idea is to define the
semantics of
effects equationally, with respect to the term model generated by
operations $\mathtt{op} \in \Sigma$, the signature of an algebra.  When
combined with the notion of an \textit{effect handler}, an
idea originally
introduced by \citet{CF94} and later investigated by \citet{PP13}, algebraic
effects generalize to more complex control effects yet still justify equational
reasoning.  The monoidal structure of algebraic effects is well
known~\cite{HPP06}; more recent work has studied the relationship between
monad transformers and modular algebraic effects~\cite{SPWJ16}.

In our setting, an event interface such as \ilc{stateE}
(Figure~\ref{fig:state}) defines an effect signature $\Sigma$, and its
constructors \ilc{Get} and \ilc{Put s} define the operations.
\citeauthor{PP13} used the notation $\mathtt{op}(x\!:\!X.\, M)$,
corresponding to the ITrees \ilc{Vis op (fun x:X => M)} construct, and called it ``operation application''.  They
axiomatized the intended semantics of effects via equations on operation
applications---for example, the fact that two \texttt{get} operations can
get collapsed into one was expressed by the equation
$\mathtt{get}(x\!:\!S.\, \mathtt{get}(y\!:\!S.\, k x y)) =
\mathtt{get}(x\!:\!S.\, k x x)$.
For ITrees, we prove such equations relative to an interpretation
of the events, as in Section~\ref{sec:state}.

The handlers of algebraic effects specify the data needed to construct an
interpretation of the effect; they have the form
$\mathtt{handler}\{\mathtt{return}~x \mapsto f(x), (\mathtt{op}(y;\kappa)
\mapsto h(y,\kappa))_{\op\in\Sigma}\}.$ In terms of our notation, the
\texttt{return} component of the handler specifies the \ilc{Ret} case of an
interpreter, and the sum over operation interpretations is written using a
dependent type.  Here, $h$ corresponds to the most general elimination
form for the ITree \ilc{Vis} constructor, which is a function of type
\ilc{forall X, E X -> (X -> itree E R) -> M R} for \ilc{M} an iterative monad.
However, Coq prevents us from creating a general-purpose interpeter
parameterized by such a type---it needs to see the definition of the handler's
body to verify the syntactic guardedness conditions.

In a language such as Eff~\cite{BP15}, \sz{cite Frank
  too?} which supports algebraic effects natively, the operational semantics
plumbs together the
continuations with the appropriate handlers, scoping them
according to the dynamic semantics of the language.  In our case, we must
explicitly invoke functions like \ilc{interp_state} as needed, possibly after
massaging the structure of events so that they have the right form.

\citet{JSV10} studied the contextual equivalences induced by interpretations of
standard effects.  Most saliently, their paper developed its theory in terms of
observations of ``computation trees,'' which are ``incompletely known''
ITrees---they are inductively defined, and hence finite, but may also include
$\bot$ leaves that denote (potential) divergence.  \citeauthor{JSV10} showed how to endow
the set of computation trees with a CPO structure based on approximation
($\bot \sqsubseteq t$ for any tree $t$) and use that notion to study contextual
equivalences induced by various interpreters.  The techniques proposed there
should be adaptable to our setting: instead of working with
observational partial orders, we might choose to work more directly with the ITree
structures themselves.

\subsection{Effects in Type Theory}

Most of the work discussed above was done either in the context of programming
languages with support for general recursion or in a theoretical ``pen and
paper'' setting, rendering these approaches fundamentally different to the ITree
library which is formalized in a total language.
Work more closely related to ITrees is that
undertaken in the context of dependent type theory.

The earliest work on
mixing effects with type theory was done by \citet{HS00}, followed by
Hancock's dissertation~\cite{hancock-thesis}.  This line of work, inspired by
monads and especially Haskell's IO monad, showed how to encode such constructs
in Martin-L\"{o}f type theory.  Those theories, in contrast to ITrees, do
not allow silent steps of computation, instead integrating guarded or sized
coinductive types as part of a strong discipline of total functional
programming.  The benefit of this is that strong bisimilarity is the only
meaningful notion of equivalence; the drawback is that they cannot handle
general recursion.  Later work on object
encodings~\cite{oo-dependent-types} did consider recursive computations, though
it did not study their equational theory or the general case of implementing
interpreters within the type theory, as we have done.  More recently,
\citet{abel2017interactive} have demonstrated the applicability of these ideas
in Agda.  Although their paper includes a proof of the correctness of a stack
object (among other examples), they do not focus on the general
equational theory of such computations.

As mentioned previously, Capretta proposed using the ``delay monad''
to encode general recursion in a type theory, as we do here, though his paper used
strong bisimulation as the notion of equivalence.  The delay monad can be seen
as either an ITree without the \ilc{Vis} constructor or, isomorphically, an
ITree
of type \ilc{itree emptyE R}.  The main theoretical contribution of that paper
was showing that the monad laws hold and that the resulting system is expressive
enough to be Turing complete.  Subsequent work explored the use of the delay
monad for defining operational semantics~\cite{danielsson2012operational} and
studied how to use quotient types \cite{CUV15} or higher inductive
types~\cite{ADK17} to define equivalence up to \ilc{Tau}, which we take
as the basis for most of our equational theory.  Because we are working in Coq,
which does not have quotient or higher inductive types, we must explicitly use
setoid rewriting, requiring us to prove that all morphisms respect the
appropriate equivalences.

\citet{mcbride-free}, building on Hancock's earlier work, used what he
called the ``general monad'' to implement effects in Agda.  His monad variant is
defined inductively as shown below.
\begin{lstlisting}[style=customcoq, basicstyle=\small\ttfamily]
Inductive General (S:Set) (T : S -> Set) (X : Set) : Set :=
| RetG (x : X)
| VisG (s:S) (k : T s -> General S T X).
\end{lstlisting}
\noindent Its interface replaces our single \inlinecoq{E : Type -> Type}
parameter with \inlinecoq{S : Type} and a type family \inlinecoq{S -> Type} to
calculate the result type of the event.  McBride proposed
encoding recursion as an (uninterpreted) effect, as we present in
Section~\ref{sec:recursion}\bcp{Can we be explicit about whether we got the
  idea from him?}\sz{I believe that we reinvented the same idea}.  In particular, he shows how to give a
semantics to recursion using first a ``fuel''-based (a.k.a. step-indexed) model
and then by translation into Capretta's delay monad.  The latter can be seen as
a version of our \ilc{interp_mrec}, but one in which all of the effects must be
handled.  Our coinductively defined interaction trees also support a general
fixpoint combinator directly, which is impossible for the \ilc{General} monad.


The FreeSpec Coq library, implemented by \citet{freespec}, uses a ``program
monad'' to model components of complex computing systems.  The program monad is
essentially an inductive version of \ilc{itree}\footnote{The original version of
  FreeSpec also included a \ilc{bind} constructor, but, following our ITrees
  development, it was removed in favor of defining bind.} (without
\ilc{Tau}).  What we call ``events,'' the FreeSpec project calls ``interfaces.''
The FreeSpec project is primarily concerned with modeling first-order, low-level
devices for which general recursion is probably not needed\bcp{Yann's talk at
  the DeepSpec workshop suggested defining it the same way we do}.  Its
library offers various composition operators, including a form of concurrent
composition, and it includes a specification logic that helps prove (and
automate proofs of) properties about the systems being modeled. However, due to
FreeSpec's use of the inductive definition, such systems must be structured as
acyclic graphs.  Nevertheless, FreeSpec doesn't eschew coinduction
altogether---as we explain below, it, like CompCert, defines the environment in
which the program runs coinductively.  FreeSpec's handlers are thus capable of
expressing diverging computations, but it does not support the equational
reasoning principles that we propose.

\subsection{Composition with the Environment}

An idea that is found in several of the works discussed above is the need
to characterize properties of the program's environment. Recall the \ilc{kill9}
program from earlier, which halts when the input is \ilc{9} but continues
otherwise.  One might wish to prove that, if the environment never supplies the
input \ilc{9}, the program goes on forever.  In a more realistic setting
like CompCert, one might wish to make assertions about externally supplied
functions, such as OS calls, \texttt{malloc} or \texttt{memcpy}, or to reason
about the accumulated output on some channel such as the terminal.

The behavior of the environment is, in a sense, \textit{dual} to the
behavior of the program.  CompCert, for example, formulates the environment
as a coinductively defined ``world,'' whose definition is (a richer version
of) the following:

\begin{lstlisting}[style=customcoq, basicstyle=\small\ttfamily]
CoInductive world : Type :=
  World (io : string -> list eventval -> option (eventval * world))
\end{lstlisting}

\noindent Here the \ilc{string} and list of \ilc{eventval}s are the
\textit{outputs} of the event (they are provided by the program), and the result
(if any) is a returned value and a new world.  The environment's state is
captured in the closure of the \ilc{io} function.  Transliterating this type to
our setting we arrive at:

\snamedsection{world}

\citet{freespec} use a definition very close to this (without the
option) to define a notion of ``semantics'' for the program monad.  Given
such a definition, one can define a world that satisfies a
certain property (for
example, one that never produces \ilc{9} as an answer) and use it to constrain
the inputs given to the program, by ``running'' the program under consideration
in the given world.  CompCert defines ``running'' via a predicate called
\ilc{possible_trace} that matches the answers provided by the \ilc{io} function
to the events of the program trace.


The CertiKOS project~\cite{cal,ccal} takes the idea of composing a program
with its environment even further.  Their Concurrent Certified Abstraction
Layers (CCAL) framework also uses a trace-based formulation of semantics.  In their
context, traces are called \textit{logs} and (concurrent) components are given
semantics in terms of sets of traces.  Each component (\textit{e.g.}, a thread) can be
separately given a specification in terms of its interface to (valid) external
environments, which encode information about the scheduler and assumptions about
other components in the context.  A layer interface can ``focus'' on subsets of
its concurrently executing components; when it is focused on a single,
sequential thread, the interface is a deterministic
function from environment
interactions (as represented by the log) to its next action.  The parallel layer
composition operation links two compatible layers by ``running'' them together
(as above) according to the schedule (inputs to one
component can be provided by outputs of the other).  In this case, one thread's
behaviors influence another thread's environment.  They formulate such
interactions in terms of concepts from game semantics, which gives rise to a
notion of refinements between layer specifications.  Layers have the
symmetric monoidal structure familiar from algebraic effects.

We conjecture that the sequential behavior of the CCAL system could be expressed
in terms of ITrees and that the concurrent composition operations of the
framework could be defined on top of that.  Our KTree combinators already offer
a rich notion of composition, including general, mutually recursive linking,
which is similar to that offered by CCAL. Moreover, we can define similar
``running'' operations directly on ITrees, rather than on
traces, coordinating multiple ITrees via an executable scheduler.  This means that, besides proving properties of the
resulting system, we can extract executable test cases~\cite{deepweb19}.

\subsection{Formal Semantics}
\label{sec:compositional-semantics}

\gmm{If we make the paper a bit more about denotational semantics, it probably makes sense to move this earlier in the related work section.}

There are a plethora of techniques used to describe the semantics of programming
languages within proof assistants.
In evaluating these techniques, we need to consider both the simplicity of the definitions and their robustness to language extensions.
The former is important because complex models are difficult to reason about, while the latter is important because seemingly small changes sometimes cascade through a language, invalidating previous work.

Denotational semantics translate the object language (\textit{e.g.}, \Imp{} or \Asm{}) into the meta-language (\textit{e.g.}, Gallina), seeking to leverage the existing power of the proof assistant.
\citet{chlipala2007} also uses denotational semantics to verify a compiler, but
in a simpler setting with a normalizing source language, and with models
individually tailored to the intermediate languages.
In contrast, ITrees serve as a common foundation for both semantics
in our case-study compiler, and an equational theory enabling the verification of a
termination-sensitive theorem.
As we saw in Section~\ref{sec:imp}, impure features such as nontermination can make this difficult, as proof assistants often\bcp{??} include only a total function space.
One way to circumvent this limitation is via a ``fuel''-based semantics, where computations are approximated to some finite amount of unwinding.
\citet{owens-functional-big-step} use this approach to develop functional big-step semantics.
To reason about nonterminating executions,
\citet{owens-functional-big-step} leverages the classical nature of the HOL logic to assert that, if no amount of fuel is sufficient for termination, then the computation diverges.
They further show how oracle semantics~\cite{hobor-oracle-semantics} can be used to enrich this language with both IO and nondeterminism.
In practice, the approach is quite similar to ITrees, except that we can
omit the fuel and instead directly construct the infinite computation tree.
With ITrees, events encode oracle queries and \ilc{Tau}s represent internal steps, which may lead to divergence.
Though, as we showed in Section~\ref{sec:case-study}, users of ITrees are mostly insulated from \ilc{Tau}s when using the combinators from Section~\ref{sec:recursion}.

The approach of \citet{owens-functional-big-step} is reminiscent of traditional step indexing~\cite{ahmed-phd}, in which the meaning of a program is described by a set of increasingly accurate approximations.
Coinductive interaction trees enable us to describe an entire, possibly infinite, computation once and for all.
Post-facto, ITrees can be easily approximated by a collection of trees or traces (Section~\ref{sec:trace}), providing a means to recover step-indexed reasoning if desired.


Formalizations of more classic domain-theoretic denotational models
exist~\cite{benton-denotational,benton-semantics}.
Unfortunately, the learning curve for this style of denotational semantics
was widely considered to be quite steep. 
The complexity of domain theoretic models prompted exploring more operational approaches to formalizing semantics~\cite{plotkin-structural-approach,plotkin-origin}.
Big-step operational semantics share a similar flavor to denotational semantics as they both connect terms directly to their meaning.
Unfortunately, interpreting big-step semantics inductively prevents them from representing divergent computations.
Some works~\cite{delaware-a-la-carte,chlipala-compiler} avoid the issue of nontermination entirely, ascribing semantics only to terminating executions.
\citet{chargueraud-pretty-big} provides a technique for avoiding the problem by duplicating the semantics both inductively and coinductively.
They\bcp{who?} argue that such duplication can be automated and therefore should not be overly burdensome.
The functional style of this ``pretty big step'' semantics is quite similar to functional denotational semantics, and thus bears a resemblance to ITrees.
ITrees avoid the need to duplicate the semantics by giving a data representation rather than a propositional representation.

\citet{leroy-coinductive} give an in-depth discussion relating inductive and coinductive semantic styles,
providing an inductive judgment for ``terminates in a value (and a trace)'' and a coinductive judgment for ``diverges (with an infinite trace).''
Relating these semantics can be difficult, and the proofs sometimes rely on
classical logic.  \bcp{How does this paragraph relate to ITrees?}

\bcp{Overall comment: The related work section at the moment is a bit
  exhausting---it feels like we don't have a clear criterion for inclusion
  or exclusion, which leads to not knowing quite what we want to say about
  some of the things beyond just explaining them, and this results in a bit
  of a hodgepodge.  I am not suggesting dropping a lot of material (which
  could be dangerous if reviewers were looking for it), but I think we could
  still tighten the presentation a bit if we try harder to explain why we
  are talking about each thing.}

\section{Conclusion}
\label{sec:conclusion}

Interaction trees are a promising basis on which to build denotational semantics
for impure and recursive computations in theorem provers like Coq.
We have established a solid theoretical foundation and demonstrated a usable
realization of the theory in Coq, as well as its practical use via a demonstrative
verified compiler example.

A natural question concerns the ease of transferring this work to other proof
assistants. While the theory does not depend on Coq, adapting the choices made
through the conception of the library to other languages for formal verification
may require some non-trivial work.

We leverage mainly three features of Coq: extraction,
coinductive types, and higher-order types.
While extraction (or, alternatively, executability) is available in most popular modern proof systems, the two
other characteristics can raise challenges in adapting this work.
More specifically, Lean~\cite{Lean} lacks coinductive types, making it seemingly
inadequate to the task.
Isabelle/HOL~\cite{Isabelle} lacks higher-order types, which appears
to be a serious obstacle to a faithful translation of our work. An expert might find a way
to encode the ITree generic event types, but we are unsure about the feasibility of the task:
this might be a stumbling block.
Finally, Agda~\cite{Agda} would be perfectly suitable, and some related work of
a similar structure, discussed in Section~\ref{sec:related-work}, actually enjoy
a formalization in this proof assistant.

Many directions for further exploration remain. Other kind of effects,
such as nondeterminism and concurrency, are instrumental in the
modeling of some systems: developing simulations and reasoning principles
for those represent a valuable challenge.
We are accumulating empirical evidence that ITrees are both expressive enough
to be adequate in various targets for formalization, while also being very convenient
to work with. Their compositional and modular nature seems to lead to
better proofs than traditional approaches, notably when it comes to reasoning
about control flow.
Building formal bridges and comparisons to
related approaches such as domain theoretic denotations, operational semantics,
step-indexed-based approaches, or game semantics models would be a great
opportunity to attempt to ground this empirical evidence.
Finally, the versatility of ITrees make them a potential fit in many contexts.
Stress-testing their viability and scalability is a major avenue we are beginning
to explore.

\subsection*{Acknowledgements}

\sz{fix this up}
This work was funded by the National Science
Foundation's Expedition in Computing \emph{The Science of Deep Specification}
under the awards 1521602 (Appel), 1521539 (Weirich, Zdancewic, Pierce)
with additional support by the NSF projects \emph{Verified
  High Performance Data Structure Implementations}, {\em Random Testing for Language Design}, award 1421243
(Pierce), ONR grant {\em REVOLVER} award N00014-17-1-2930,
and by the Basic Science Research Program through the National Research Foundation of Korea (NRF) funded by the Ministry of Science and ICT (2017R1A2B2007512).
We are especially thankful to Joachim Breitner and Dmitri Garbuzov for
early contributions to this work.
We are grateful to all the members of the DeepSpec project for
their collaboration and feedback, and
we greatly appreciate the the reviewers' comments and suggestions.

\balance
\bibliography{ref}

\end{document}